\documentclass[british,useAMS,usenatbib]{mn2e}
\usepackage[T1]{fontenc}
\usepackage[latin9]{inputenc}
\usepackage{xcolor}
\usepackage{array}
\usepackage{booktabs}
\usepackage{units}
\usepackage{amsmath}
\usepackage{amssymb}
\usepackage{wasysym}
\usepackage{graphicx}
\usepackage{esint}
\usepackage[authoryear]{natbib}
\PassOptionsToPackage{normalem}{ulem}
\usepackage{ulem}
\usepackage{subscript}

\makeatletter

\providecommand{\tabularnewline}{\\}
\newcommand{\lyxdot}{.}

\providecolor{lyxadded}{rgb}{0,0,1}
\providecolor{lyxdeleted}{rgb}{1,0,0}
\usepackage[caption=false]{subfig}
\usepackage{savesym}
\savesymbol{iint}
\savesymbol{iiint}
\savesymbol{iiiint}
\savesymbol{longtable}
\usepackage{amsmath}
\restoresymbol{}{iint}
\restoresymbol{}{iiint}
\restoresymbol{}{iiiint}
\restoresymbol{}{longtable}
\savesymbol{bibhang}
\usepackage[authoryear]{natbib}
\restoresymbol{}{bibhang}
\setcitestyle{aysep={}}

\usepackage{tablefootnote}

\makeatother

\usepackage{babel}
\begin{document}
\title[Testing the Modern Merger Hypothesis]{Testing the Modern Merger Hypothesis via the Assembly of Massive Blue Elliptical Galaxies in the Local Universe}
\author[T. Haines et al.]{Tim Haines,$^{1,2}$
\thanks{E-mail:thaines@astro.wisc.edu}
\thanks{Visiting Astronomer, Kitt Peak National Observatory, National Optical Astronomy Observatory, which is operated by the Association of Universities for Research in Astronomy, Inc. (AURA) under cooperative agreement with the National Science Foundation.}
D. H. McIntosh,$^{2}$ S. F. S\'anchez,$^{3}$ C. Tremonti,$^{1}$ and G. Rudnick$^{5}$\\
$^{1}$Department of Astronomy, University of Wisconsin-Madison, 475 N. Charter Street, Madison WI 53706-1582, USA\\
$^{2}$Department of Physics and Astronomy, University of Missouri - Kansas City, 5110 Rockhill Road, Kansas City, MO 64110, USA\\
$^{3}$Instituto de Astronom\'\i a,Universidad Nacional Auton\'oma de Mexico, A.P. 70-264, 04510, M\'exico,D.F.\\
$^{5}$Department of Physics and Astronomy, The University of Kansas, Malott room 1082, 1251 Wescoe Hall Drive, Lawrence, KS 66045, USA}
\maketitle
\begin{abstract}
The modern merger hypothesis offers a method of forming a new elliptical
galaxy through merging two equal-mass, gas-rich disk galaxies fuelling
a nuclear starburst followed by efficient quenching and dynamical
stabilization. A key prediction of this scenario is a central concentration
of young stars during the brief phase of morphological transformation
from highly-disturbed remnant to new elliptical galaxy. To test this
aspect of the merger hypothesis, we use integral field spectroscopy
to track the stellar Balmer absorption and 4000\AA\  break strength
indices as a function of galactic radius for 12 massive (${\rm M_{*}}\ge10^{10}{\rm M_{\odot}}$),
nearby (${\rm z}\le0.03$), visually-selected plausible new ellipticals
with blue-cloud optical colours and varying degrees of morphological
peculiarities. We find that these index values and their radial dependence
correlate with specific morphological features such that the most
disturbed galaxies have the smallest 4000\AA\  break strengths and
the largest Balmer absorption values. Overall, two-thirds of our sample
are inconsistent with the predictions of the modern merger hypothesis.
Of these eight, half exhibit signatures consistent with recent minor
merger interactions. The other half have star formation histories
similar to local, quiescent early-type galaxies. Of the remaining
four galaxies, three have the strong morphological disturbances and
star-forming optical colours consistent with being remnants of recent,
gas-rich major mergers, but exhibit a weak, central burst consistent
with forming $\sim5\%$ of their stars. The final galaxy possesses
spectroscopic signatures of a strong, centrally-concentrated starburst
and quiescent core optical colours indicative of recent quenching
(i.e., a post-starburst signature) as prescribed by the modern merger
hypothesis.
\end{abstract}

\section{Introduction}

\let\thefootnote\relax\footnote{The WIYN Observatory is a joint facility of the University of Wisconsin-Madison, Indiana University, the National Optical Astronomy Observatory and the University of Missouri.}The
stellar growth of high-mass ($>10^{10}{\rm \: M}_{\odot}{\rm h}^{-2}$),
quiescent (non-star-forming) galaxies remains an important topic in
galaxy evolution studies. Large and deep redshift surveys have conclusively
shown that these red galaxies have grown significantly since at least
redshift $z\sim2-3$ in terms of both their number density \citep[e.g.,][]{bell2004,brown2007,martin2007,brammer2009}
and average sizes at a fixed stellar mass \citep{khochfar2006,trujillo2006,trujillo2007,lee2013}.
The observed build up of this population and its conserved preponderance
of early-type galaxy (ETG) morphologies at different epochs \citep{bell20042,blanton2006,bell2012}
requires a process (or sequence of processes) which both increases
the number of galaxies with early-type morphologies \textit{and} quenches
star formation. The currently accepted explanation for this phenomenon
is the so-called blue-to-red migration \citep{faber2007,eliche-moral2010,gonccalves2012quenching}
which posits that the transformation of star-forming (blue) disk galaxies
into quiescent, red ETGs is responsible for the increasing population
of red sequence galaxies over cosmic time. One possible, but well-accepted,
mechanism driving the blue-to-red migration is the modern merger hypothesis
\citep{hopkins2008}. Under this model, a new massive quiescent elliptical
galaxy is built by transforming two star-forming disk galaxies into
one concentrated quiescent red ETG via a gas-rich major (i.e., nearly
equal mass) merger.

Yet, in detail, the red ETG population contains myriad morphological
types in detail from early-type spiral disks, to lenticular (S0) galaxies,
to the well-known dichotomy of ellipticals \citep[e.g.,][]{kormendy2009},
making disentangling which migration processes are dominant a daunting
task that likely depends on galaxy morphology, mass, environment,
and redshift. We have made progress in the local universe by focusing
on the redward migration of pure-spheroid elliptical galaxies \citep[hereafter, M14]{mcintosh2014anew}.
This choice is motivated by the fact that the formation of new ellipticals
is theoretically tied to a \textit{single} mechanism that both transforms
disks \citep[e.g., major merging,][and references therein]{toomre1972,barnes1992}
and is hierarchically motivated \citep{white1978,kauffmann1993}.

Simulations of gas-rich major mergers provide a key prediction which
allows us to qualitatively identify major merger remnants. During
the merging process, gravitational torques deposit cold gas in the
centre of the remnant and produce a centrally-concentrated burst of
star formation \citep{barnes1991,barnes1996,mihos1996}. According
to the modern merger hypothesis, this is quickly followed by the ignition
of an active galactic nucleus (AGN) which ultimately quenches star
formation \citep{springel2005,schawinski2007,schawinski2010} and
forms a quiescent elliptical galaxy \citep{hopkins2006,hopkins20082}.
This is supported by the bluer colours of early-type galaxies hosting
an AGN, compared with their non-active counterparts \citep{sanchez2004colorsof}.
The merging process finally produces a relaxed remnant with the structure
and kinematics of observed moderate-mass spheroidal galaxies \citep{cox2006,naab2006}.

We, therefore, anticipate ellipticals which are plausible remnants
of recent gas-rich mergers to have bluer colours as compared to their
``normal'' counterparts on the colour-mass relation due to recent,
enhanced star formation. Such blue ellipticals are rare. For example,
\citet{fukugita2004} found 2 of the 210 ellipticals in their sample
of 1600 bright galaxies from the SDSS at z<0.12 to be blue ellipticals,
M14 found 1602 blue ellipticals in their complete sample of 63,454
massive $\left({\rm M_{*}}>10^{10}{\rm M_{\odot}h^{-2}}\right)$ galaxies
at z<0.08, and \citet{lee2006} found 29 in their sample of 1949 galaxies
with spectroscopic redshifts $\lesssim0.4$ in the GOODS fields. Despite
their rarity, these galaxies are important as they build the transition
from star-forming disks in the blue cloud to massive, quiescent ellipticals
on the red sequence \citep{bell2007}. The robust sample of blue elliptical
galaxies seen in M14 provides a good starting point for testing methods
to distinguish recent elliptical mass assembly.

Over the last decade, spatially-resolved spectroscopic surveys such
as SAURON \citep{bacon2001thesauron} and ATLAS3D \citep{cappellari2011theatlas3d}
have drastically enhanced our understanding of the content and structure
of cluster and field ellipticals. Likewise, projects such as PINGS
\citep{rosales-ortega2010pingsthe}, TYPHOON \citep{sturch2012typhoon},
and the CALIFA survey \citep{sanchez2012} have provided an unprecedented
level of detail in spiral galaxies. Aside from serendipitous discoveries
\citep[e.g.,][]{fukugita2004}, little work has been brought forth
from the population of blue elliptical galaxies. Further, previous
studies of these galaxies have been limited to evaluating only the
central regions \citep[e.g.,][]{schawinski2009,tojeiro2013thedifferent}
which provides a narrow perspective of the mass assembly history.
In this work, we propose to analyse the 4000\AA  break strength and
Balmer absorption line indices out to $\sim3$ half-light radii (coverage
previously only achieved for red elliptical galaxies, e.g. \citealt{weijmans2009})
to examine the spatial star formation histories of this rare, but
important population of elliptical galaxies in the local universe.

Our paper is organized as follows. Section 2 details our sample selection,
observation setups, and data reduction pipeline. Section 3 outlines
our model construction. In Section 4, we provide results from our
comparison of the spatially-resolved spectra to a suite of star formation
history (SFH) models. Section 5 contains the discussion of our results.
Finally, in Section 6 we provide a summary and conclusions. Throughout
this paper we calculate comoving distances in the $\Lambda$CDM concordance
cosmology with $\Omega_{{\rm m}}=0.3$, $\Omega_{\Lambda}=0.7$, and
assume a Hubble constant%
\footnote{The stellar masses from M14 are calculated assuming ${\rm H_{0}}=100h$.%
} of ${\rm H_{0}}=70\mathrm{\; km\; s^{-1}\; Mpc^{-1}}$. Magnitudes
from the Sloan Digital Sky Survey \citep[SDSS]{york20002} are on
the AB system such that $m_{{\rm AB}}=m+\Delta m$ where $\Delta m=(-0.036,+0.012,+0.010,+0.028,+0.040)$
for $(u,g,r,i,z)$ \citep{yang2007galaxygroups}.

\section{Data Sample and Observations}

\subsection{Sample Selection}

\label{sub:sampleSelection}
\begin{figure}
\includegraphics{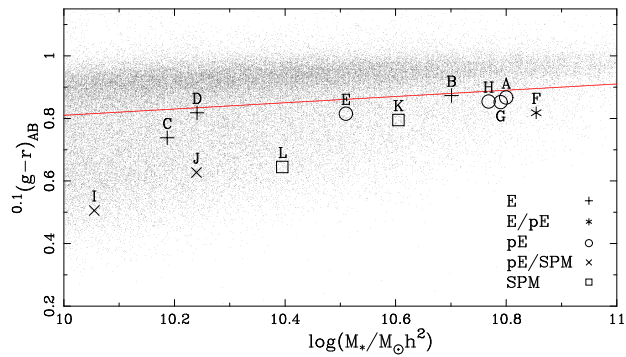}

\protect\caption{\label{fig:sampleSelection_colorMass}Colour-mass diagram for the
63454 nearby ($0.01\le z\le0.08$), high-mass ($M_{{\rm *}}\ge10^{10}M_{\odot}h^{-2}$)
SDSS DR4 galaxies from M14 with our sample of 12 blue ellipticals
overlaid. The red line shows the M14 colour cut, symbols represent
visual classifications from M14. Refer to Table \ref{tab:sample}
for letter labels.}
\end{figure}

\textcolor{black}{To test the prediction
of the modern merger hypothesis that gas-rich major mergers lead to
the construction of massive ellipticals with temporarily blue colours,
we acquire observationally-intensive IFU spectroscopy using long exposure
times to obtain spatially-resolved spectra beyond the bright central
bulge.} We then use the ${\rm H}\delta_{{\rm A}}$ and ${\rm H}\gamma_{{\rm A}}$
Lick indices and the narrow-band 4000\AA  break to map the recent
star formation histories as a function of galactic radius for a collection
of visually-classified%
\footnote{\textcolor{black}{The morphological classifications of M14 were determined
by manual inspection of the images by people. Here and throughout
the paper, we use the phrase ``visually-classified'' to mean classified
``by eye.''}%
} plausible new ellipticals provided in M14. Using the New York University
Value-Added Galaxy Catalogue \citep[NYU-VAGC,][]{blanton2005} based
on the SDSS Data Release 4 \citep[DR4,][]{adelman-mccarthy2006},
M14 construct a complete sample of nearby ($0.01\le z\le0.08$), high-mass
($M_{{\rm *}}\ge10^{10}M_{\odot}h^{-2}$) galaxies. Using the rest
frame $(g-r)$ Petrosian colours, the absolute r-band magnitudes,
and the \citet{bell2003} $M/L$ prescription to measure stellar mass,
they apply the empirical colour cut $^{0.1}(g-r)\le0.81+0.1\left[\log_{10}({\rm M}_{{\rm gal,}{\star}}/{\rm M}_{\odot}{\rm h}^{-2})-10.0\right]$
to select blue-cloud galaxies (see Fig. \ref{fig:sampleSelection_colorMass})
where the colors have been k-corrected to $z=0.1$. They isolate early-type
(spheroid-dominated) galaxies using the common $r$-band concentration
cut of ${\rm R_{90}/{\rm R_{50}}}\geq2.6$ \citep{strateva2001},
where ${\rm R_{90}}$ (${\rm R_{50}}$) is the radius inside which
90\% (50\%) of the light is contained to produce a set of 8403 unusually
blue ETGs.

M14 then use visual classification with careful morphological inspection
to identify plausible new merger remnants from this set of blue ETGs.
Visual inspection further allows them to remove contamination by bulge-dominated
spirals (S) and inclined disks (iD), plus galaxies with uncertain
morphology (U), and galaxies affected by artefacts. Their final catalogue
consists of 1602 massive, blue elliptical (E,85\%), peculiar elliptical
(pE,8\%), and spheroidal post-merger (SPM,7\%) galaxies with high
classifier-to-classifier agreement, defined as a minimum of three
out of four classifiers in accord. The details of these identifications
and their robustness are described in detail in M14. Briefly, the
E, pE and SPM types represent three morphological bins that plausibly
span a \textit{qualitative} time sequence since merging (assuming
that all blue elliptical galaxies are derived from gas-rich major
merging)\textcolor{black}{. At one extreme (E), galaxies appear to
be dynamically relaxed ellipticals with little or no evidence of recent
tidal activity, contrasted by those at the other extreme which appear
to be freshly coalesced with very disturbed morphologies (SPM). In
between the two extremes (pE), we find galaxies which appear to be
relaxed ellipticals with either a modest external morphological peculiarity
(e.g., a loop or fan feature) or an internal structure like a dust
lane. We parametrize the level of visual tidal features by defining
a plausible post-merge}r type, T\textsubscript{ppm}, for each blue
ETG based on the sum of its four classifications weighted as follows:
\begin{align}
{\rm T_{ppm}=}\sum_{i=1}^{4}w_{c,i}\label{eq:tppm}
\end{align}
where $w_{c,i}$ is the ${\rm i^{th}}$ classification such that $w_{c}$
= -1 (S/iD), -0.5 (U), 1 (E), 2 (pE), and 3 (SPM); e.g., T\textsubscript{ppm}
= 8 for four-way pE agreement. We use negative weights to distinguish
definite disks and uncertain classifications from (positive $w_{c}$
) plausible post-merger systems\textcolor{black}{. We investigate
a subset of the 8403 blue ETGs from M14 using a relaxed classifier
agreement such that ${\rm T_{ppm}\ge2.0}$ to produce a sample of
1915 E (71\%), pE (6\%), SPM (6\%), plus 17\% with combinations of
E+pE+SPM classifications (i.e., classifiers agree that these objects
are not S or iD despite disagreement over specific plausible post-merger
type). For example, nyu100917 appears to be a relatively smooth elliptical
with a large loop structure (see Figure \ref{fig:masterMosaic}).
The classifiers in M14 were evenly split between the pE and SPM classification:
falling below the 75\% classifier agreement requirement. Using our
relaxed classifier allows this galaxy to be included in our sample.}

\textcolor{black}{To maximize spatial coverage of the spectra, we
require each galaxy to fill the FOV of its resp}ective spectrograph
(each spectrograph used in this work is$\sim70\arcsec$ on a side)
in one of two ways: (i) twice the galaxy's ${\rm R}90\simeq70\arcsec$
or (ii) the galaxy has loops, arms, or other structures which extend
beyond its main body to fill the FOV. These requirements reduce the
range of redshifts for our targets to $z\lesssim0.03$ and provide
a sample of 111 E (57\%), pE (11\%), SPM (10\%), and other (${\rm T_{ppm}}\ge2$
, 23\%) galaxies. Based on observational availability, we select a
set of 12 (3 E, 3 pE, 2 SPM, and 4 other) which nearly equally sample
the morphology space while also selecting interesting candidates for
unique star formation histories based on a range of morphological
details such as circumnuclear blue rings or large dust features. In
Table \ref{tab:sample}, we list physical characteristics and morphological
classifications for our sample. 
\begin{table*}
\protect\caption{Sample\label{tab:sample}}
\begin{tabular}{>{\raggedright}p{0.8in}r@{\extracolsep{0pt}.}lr@{\extracolsep{0pt}.}lr@{\extracolsep{0pt}.}lr@{\extracolsep{0pt}.}lr@{\extracolsep{0pt}.}lr@{\extracolsep{0pt}.}lr@{\extracolsep{0pt}.}lr@{\extracolsep{0pt}.}lcc}
\addlinespace[-0.6em]
\multicolumn{19}{r}{}\tabularnewline
\midrule
\midrule 
{\small{}~~~Name} & \multicolumn{2}{c}{{\small{}z\textsubscript{published}}} & \multicolumn{2}{c}{{\small{}R50}} & \multicolumn{2}{c}{{\small{}c\textsubscript{r}}} & \multicolumn{2}{c}{{\small{}g}} & \multicolumn{2}{c}{{\small{}${\rm M_{r}}$}} & \multicolumn{2}{c}{{\small{}M\textsubscript{{*}}}} & \multicolumn{2}{c}{{\small{}Classification\dag{}}} & \multicolumn{2}{c}{{\small{}Agreement}} & {\small{}T\textsubscript{{\small{}ppm}}} & ID\tabularnewline
{\small{}~~~~~(1)} & \multicolumn{2}{c}{{\small{}(2)}} & \multicolumn{2}{c}{{\small{}(3)}} & \multicolumn{2}{c}{{\small{}(4)}} & \multicolumn{2}{c}{{\small{}(5)}} & \multicolumn{2}{c}{(6)} & \multicolumn{2}{c}{{\small{}(7)}} & \multicolumn{2}{c}{{\small{}(8)}} & \multicolumn{2}{c}{{\small{}(9)}} & {\small{}(10)} & (11)\tabularnewline
\midrule
{\small{}nyu100917} & {\small{}0}&{\small{}027} & {\small{}2}&{\small{}9} & {\small{}3}&{\small{}6} & {\small{}15}&{\small{}10} & {\small{}-21}&{\small{}11} & {\small{}10}&{\small{}24} & \multicolumn{2}{c}{{\small{}pE/SPM}} & \multicolumn{2}{c}{{\small{}50}} & {\small{}10} & J\tabularnewline
{\small{}nyu22318} & {\small{}0}&{\small{}030} & {\small{}5}&{\small{}9} & {\small{}3}&{\small{}0} & {\small{}14}&{\small{}79} & {\small{}-21}&{\small{}72} & {\small{}10}&{\small{}79} & \multicolumn{2}{c}{{\small{}pE}} & \multicolumn{2}{c}{{\small{}75}} & {\small{}7} & G\tabularnewline
{\small{}nyu535845} & {\small{}0}&{\small{}014} & {\small{}6}&{\small{}2} & {\small{}3}&{\small{}0} & {\small{}13}&{\small{}99} & {\small{}-20}&{\small{}81} & {\small{}10}&{\small{}24} & \multicolumn{2}{c}{{\small{}E}} & \multicolumn{2}{c}{{\small{}75}} & {\small{}5} & D\tabularnewline
{\small{}nyu541044} & {\small{}0}&{\small{}020$^{(a)}$} & {\small{}6}&{\small{}0} & {\small{}3}&{\small{}5} & {\small{}13}&{\small{}55} & {\small{}-22}&{\small{}24} & {\small{}10}&{\small{}61} & \multicolumn{2}{c}{{\small{}SPM}} & \multicolumn{2}{c}{{\small{}75}} & {\small{}11} & K\tabularnewline
{\small{}nyu598180} & {\small{}0}&{\small{}018} & {\small{}4}&{\small{}4} & {\small{}2}&{\small{}9} & {\small{}14}&{\small{}58} & {\small{}-20}&{\small{}75} & {\small{}10}&{\small{}19} & \multicolumn{2}{c}{{\small{}E}} & \multicolumn{2}{c}{{\small{}75}} & {\small{}5} & C\tabularnewline
{\small{}nyu619567} & {\small{}0}&{\small{}029} & {\small{}5}&{\small{}2} & {\small{}3}&{\small{}1} & {\small{}14}&{\small{}16} & {\small{}-22}&{\small{}31} & {\small{}10}&{\small{}85} & \multicolumn{2}{c}{{\small{}E/pE}} & \multicolumn{2}{c}{{\small{}50}} & {\small{}6} & F\tabularnewline
{\small{}nyu674881} & {\small{}0}&{\small{}022} & {\small{}8}&{\small{}3} & {\small{}3}&{\small{}3} & {\small{}13}&{\small{}79} & {\small{}-22}&{\small{}10} & {\small{}10}&{\small{}80} & \multicolumn{2}{c}{{\small{}pE/U}} & \multicolumn{2}{c}{{\small{}50}} & {\small{}3} & A\tabularnewline
{\small{}nyu719486} & {\small{}0}&{\small{}028} & {\small{}2}&{\small{}4} & {\small{}3}&{\small{}3} & {\small{}15}&{\small{}08} & {\small{}-20}&{\small{}85} & {\small{}10}&{\small{}06} & \multicolumn{2}{c}{{\small{}pE/SPM}} & \multicolumn{2}{c}{{\small{}50}} & {\small{}10} & I\tabularnewline
{\small{}nyu742039} & {\small{}0}&{\small{}025} & {\small{}5}&{\small{}2} & {\small{}3}&{\small{}1} & {\small{}14}&{\small{}16} & {\small{}-21}&{\small{}90} & {\small{}10}&{\small{}70} & \multicolumn{2}{c}{{\small{}E}} & \multicolumn{2}{c}{{\small{}100}} & {\small{}4} & B\tabularnewline
{\small{}nyu835691} & {\small{}0}&{\small{}026$^{(b)}$} & {\small{}5}&{\small{}3} & {\small{}3}&{\small{}4} & {\small{}14}&{\small{}42} & {\small{}-21}&{\small{}41} & {\small{}10}&{\small{}40} & \multicolumn{2}{c}{{\small{}SPM}} & \multicolumn{2}{c}{{\small{}75}} & {\small{}11} & L\tabularnewline
{\small{}nyu916578} & {\small{}0}&{\small{}029} & {\small{}5}&{\small{}1} & {\small{}3}&{\small{}4} & {\small{}14}&{\small{}85} & {\small{}-21}&{\small{}34} & {\small{}10}&{\small{}51} & \multicolumn{2}{c}{{\small{}pE}} & \multicolumn{2}{c}{{\small{}75}} & {\small{}5.5} & E\tabularnewline
{\small{}nyu916757} & {\small{}0}&{\small{}028 } & {\small{}6}&{\small{}3} & {\small{}3}&{\small{}1} & {\small{}14}&{\small{}19} & {\small{}-22}&{\small{}11} & {\small{}10}&{\small{}77} & \multicolumn{2}{c}{{\small{}pE}} & \multicolumn{2}{c}{{\small{}75}} & {\small{}7} & H\tabularnewline
\bottomrule
\end{tabular}\\
N\textit{\small{}ote:}{\small{} For each galaxy in our sample, we
list (1) the ID provided in the NYU-VAGC from the SDSS DR4, (2) the
spectroscopic redshift published in SDSS or NED (objects with a superscript
are from NED), (3) the SDSS r-band Petrosian half-light radius in
arcseconds, (4) the concentration index given by the ratio of the
SDSS r-band Petrosian R90 (the radius inside which ninety percent
of the light is contained) and the half-light radius (${\rm R_{90}/{\rm R_{50}}}$),
(5) the SDSS g-band apparent magnitude, (6) the extinction- and k-corrected
(to ${\rm z=0}$ absolute SDSS r-band magnitude, (7) the stellar mass
in units of $\log\left(M_{\odot}\, h^{-2}\right)$, (8) morphological
classification (see text for details), (9) classifier agreement percentage,
(10) the plausible post-merger type parameter (Equation \ref{eq:tppm}),
and (11) the ID used to refer to each galaxy in figures ordered by
increasing ${\rm T_{ppm}}$.}\\
{\small{}\dag{} }{\footnotesize{}All classifications are given where
(9) is less than 75 per-cent.}{\small{}}\\
{\small{}(a) \citet{rothberg2006} , (b) \citet{falco1999}}
\end{table*}

In Fig. \ref{fig:sampleSelection_colorMass}, we show our sample in
the colour-mass plane using the SDSS (\textit{g-r}) colours k-corrected
to ${\rm z=0.1}$ in the AB magnitude system and the log of the stellar
mass in units of ${\rm M_{\odot}h^{-2}}$. With the exception of the
two low-mass ellipticals with smooth morphologies, we find that the
low-mass galaxies tend to have the strongest morphological peculiarities
and blue colours consistent with star-forming galaxies. The more massive
galaxies tend to exhibit green valley colours and smoother morphologies.
We note that all of the peculiar ellipticals have masses between $10^{10.5}$
and $10^{11}{\rm M_{\odot}h^{-2}}$, above which extremely few blue
galaxies lie (e.g., \citealt{baldry2004}).

\subsection{Observations}

\label{sub:observations}Our final sample of 12 plausible merger remnants
were observed over three separate runs between January 2010 and April
2011. We require moderate resolution (R$\simeq$1000) spatially-resolved
spectra with wavelength coverage spanning the 4000\AA\, break and
Balmer ${\rm H}_{\delta}$ and ${\rm H}_{\gamma}$ absorption indices
to obtain SFHs as a function of galactic radius to differentiate between
remnants of major mergers from those of other mass assembly mechanisms.
The optimal choice for large-field, spatially-resolved spectroscopy
is an integral field unit (IFU) which additionally provides concurrent
sky flux measurement to facilitate removal of contaminating sky lines
during reduction.

The first observing run was performed between January 18\textsuperscript{th}
and January 20\textsuperscript{th}, 2010 at the 3.5-meter Calar Alto
observatory in Spain using the Potsdam MultiAperture Spectrophotometer
\citep[PMAS,][]{roth2005} and the PPAK IFU \citep{kelz2006}. We
used the V1200 grating with first order and forward blaze angle centred
on 4263\AA, providing a spectral resolution FWHM of 2.7\AA, and
an observed wavelength range of 3877\AA\ to 4633\AA. The spectra
were binned 2x2, giving an instrumental resolution of 0.75\AA/pixel.
The $2.7\arcsec$ fibres provide a spatial resolution of 1.63~kpc
at z=0.03\textcolor{black}{. Three thirty-minute exposures were taken
of each of the three targets, but nyu541044 was shorted by 5 minutes
and nyu835691 by 13 minutes to reduce background during twilight (see
Table \ref{tab:Observing-Log}). }The second observing run was performed
between January 31\textsuperscript{st} and February 2\textsuperscript{nd},
2011 at the 3.5-meter WIYN telescope at the Kitt Peak National Observatory
in Arizona using the WIYNBench spectrograph and the SparsePak IFU
\citep{bershady2004}. We used the 600 lines/mm grating with a blaze
angle of $10.1^{\circ}$ centred on 5500\AA, providing a spectral
resolution FWHM of 5\AA, and wavelength range of 3577\AA\ to 6435\AA.
The spectra were binned 3x3, giving an instrumental resolution of
1.4\AA/pixel. The $4.7\arcsec$ fibres \textcolor{black}{provide
a spatial resolution of 2.82~kpc at z=0.03. Five thirty-minute exposures
were taken of each of the eight targets with ten and fifteen minutes
additional exposure to compensate for reduced transparency for nyu598180
and nyu916757, respectively. The third run was performed as part of
the Calar Alto Legacy Integral Field Area (CALIFA) survey on April
}11, 2011 using the Calar Alto telescope and PMAS with the PPAK IFU.
The survey uses a three-pointing dithering with three exposures of
700s each per dithering with the same V1200 grating setup used on
the previous PMAS observations. In Table \ref{tab:Observing-Log},
we list relevant details from our observing log.
\begin{table*}
\protect\caption{Observing Log\label{tab:Observing-Log}}
{\small{}}%
\begin{tabular}{llr@{\extracolsep{0pt}.}lr@{\extracolsep{0pt}.}llr@{\extracolsep{0pt},}lccc}
\addlinespace[-0.6em]
 &  & \multicolumn{2}{c}{} & \multicolumn{2}{c}{} &  & \multicolumn{2}{c}{} &  &  & \tabularnewline
\midrule
\midrule 
{\footnotesize{}~~~~Name} & {\footnotesize{}~~~NED Name} & \multicolumn{2}{c}{{\footnotesize{}RA}} & \multicolumn{2}{c}{{\footnotesize{}DEC}} & {\footnotesize{}~~~Source} & \multicolumn{2}{c}{{\footnotesize{}~~~~~~~Date}} & {\footnotesize{}AM} & {\footnotesize{}t\textsubscript{{\footnotesize{}exp}}} & {\footnotesize{}A\textsubscript{g}}\tabularnewline
{\footnotesize{}~~~~~(1)} & {\footnotesize{}~~~~~~~~(2)} & \multicolumn{2}{c}{{\footnotesize{}(3)}} & \multicolumn{2}{c}{{\footnotesize{}(4)}} & {\footnotesize{}~~~~~~(5)} & \multicolumn{2}{c}{{\footnotesize{}~~~~~~~~(6)}} & {\footnotesize{}(7)} & {\footnotesize{}(8)} & {\footnotesize{}(9)}\tabularnewline
\midrule
{\footnotesize{}nyu100917} & {\footnotesize{}MRK 0366} & {\footnotesize{}32}&{\footnotesize{}8898} & {\footnotesize{}13}&{\footnotesize{}9171} & {\footnotesize{}SparsePak} & {\footnotesize{}Jan 31}&{\footnotesize{} 2011} & {\footnotesize{}1.36} & {\footnotesize{}9000} & {\footnotesize{}1.26}\tabularnewline
{\footnotesize{}nyu22318} & {\footnotesize{}NGC 1149} & {\footnotesize{}44}&{\footnotesize{}3495} & {\footnotesize{}-0}&{\footnotesize{}3094} & {\footnotesize{}SparsePak} & {\footnotesize{}Feb 2}&{\footnotesize{} 2011} & {\footnotesize{}1.33} & {\footnotesize{}9000} & {\footnotesize{}1.05}\tabularnewline
{\footnotesize{}nyu535845} & {\footnotesize{}UGC 05026} & {\footnotesize{}141}&{\footnotesize{}6659} & {\footnotesize{}45}&{\footnotesize{}8472} & {\footnotesize{}PMAS} & {\footnotesize{}Jan 20}&{\footnotesize{} 2010} & {\footnotesize{}1.08} & {\footnotesize{}5400} & {\footnotesize{}0.40}\tabularnewline
{\footnotesize{}nyu541044} & {\footnotesize{}NGC 3921} & {\footnotesize{}177}&{\footnotesize{}7786} & {\footnotesize{}55}&{\footnotesize{}0788} & {\footnotesize{}PMAS} & {\footnotesize{}Jan 18}&{\footnotesize{} 2010} & {\footnotesize{}1.05} & {\footnotesize{}5100} & {\footnotesize{}0.20}\tabularnewline
{\footnotesize{}nyu598180} & {\footnotesize{}CGCG 149-037} & {\footnotesize{}127}&{\footnotesize{}3750} & {\footnotesize{}31}&{\footnotesize{}6755} & {\footnotesize{}SparsePak} & {\footnotesize{}Feb 2}&{\footnotesize{} 2011} & {\footnotesize{}1.05} & {\footnotesize{}9600} & {\footnotesize{}0.68}\tabularnewline
{\footnotesize{}nyu619567} & {\footnotesize{}IC 0669} & {\footnotesize{}166}&{\footnotesize{}8190} & {\footnotesize{}6}&{\footnotesize{}3025} & {\footnotesize{}SparsePak} & {\footnotesize{}Feb 1}&{\footnotesize{} 2011} & {\footnotesize{}1.15} & {\footnotesize{}9000} & {\footnotesize{}0.58}\tabularnewline
{\footnotesize{}nyu674881} & {\footnotesize{}NGC 6314} & {\footnotesize{}258}&{\footnotesize{}1613} & {\footnotesize{}23}&{\footnotesize{}2701} & {\footnotesize{}CALIFA} & {\footnotesize{}Apr 5}&{\footnotesize{} 2011} & {\footnotesize{}1.09} & {\footnotesize{}6300} & {\footnotesize{}0.93}\tabularnewline
{\footnotesize{}nyu719486} & {\footnotesize{}MRK 0385} & {\footnotesize{}120}&{\footnotesize{}8669} & {\footnotesize{}25}&{\footnotesize{}1027} & {\footnotesize{}SparsePak} & {\footnotesize{}Jan 31}&{\footnotesize{} 2011} & {\footnotesize{}1.03} & {\footnotesize{}9000} & {\footnotesize{}0.44}\tabularnewline
{\footnotesize{}nyu742039} & {\footnotesize{}UGC 06227} & {\footnotesize{}167}&{\footnotesize{}8640} & {\footnotesize{}47}&{\footnotesize{}0355} & {\footnotesize{}SparsePak} & {\footnotesize{}Feb 1}&{\footnotesize{} 2011} & {\footnotesize{}1.51} & {\footnotesize{}9000} & {\footnotesize{}0.20}\tabularnewline
{\footnotesize{}nyu835691} & {\footnotesize{}UGC 07560} & {\footnotesize{}186}&{\footnotesize{}7364} & {\footnotesize{}48}&{\footnotesize{}2770} & {\footnotesize{}PMAS} & {\footnotesize{}Jan 20}&{\footnotesize{} 2010} & {\footnotesize{}1.02} & {\footnotesize{}4600} & {\footnotesize{}0.19}\tabularnewline
{\footnotesize{}nyu916578} & {\footnotesize{}CGCG 216-016} & {\footnotesize{}189}&{\footnotesize{}5908} & {\footnotesize{}42}&{\footnotesize{}2055} & {\footnotesize{}SparsePak} & {\footnotesize{}Jan 31}&{\footnotesize{} 2011} & {\footnotesize{}1.11} & {\footnotesize{}9000} & {\footnotesize{}0.25}\tabularnewline
{\footnotesize{}nyu916757} & {\footnotesize{}NGC 4985} & {\footnotesize{}197}&{\footnotesize{}0504} & {\footnotesize{}41}&{\footnotesize{}6763} & {\footnotesize{}SparsePak} & {\footnotesize{}Jan 31}&{\footnotesize{} 2011} & {\footnotesize{}1.02} & {\footnotesize{}9900} & {\footnotesize{}0.29}\tabularnewline
\bottomrule
\end{tabular}{\small{}}\\
\textit{\small{}Note}{\small{}: For each galaxy in our sample, we
list (1) the ID provided in the NYU-VAGC from the SDSS DR4, (2) the
most common associated name in the NASA Extragalactic Database, (3)
the right ascension in degrees, (4) the declination in degrees, (5)
the IFU on which that data were acquired, (6) the date of observation,
(7) the average airmass of all exposures, (8) the total exposure time
in seconds, and (9) the SDSS }\textit{\small{}g-band}{\small{} foreground
extinction in magnitudes.}
\end{table*}

\subsection{Data Reduction}

We use the R3D reduction software \citep{sanchez2006} to produce
a data cube of wavelength- and flux-calibrated spectra for each galaxy,
which we then use to construct annular spectra for determining the
SFHs as a function of galactic radius. The reduction of IFU data starts
with the standard procedures for single-slit spectra, namely subtracting
the dark current and the bias. We perform cosmic ray rejection using
the method of \citet{vandokkum2001}. Because the IFU generates many
spectra (82 for SparsePak and 382 for PMAS), the remaining procedures
diverge from the normal methods of single-slit reduction. Readers
interested in the technical details of IFU data reduction are encouraged
to read \citet{sanchez2006}. Here, we provide an outline of our data
reduction pipeline.

\subsubsection{Reduction of Fibre Data}

\paragraph{Locating Spectra on the Detector}

Each exposure from the IFU produces a data frame consisting of the
dispersion axis along the horizontal, and N fibre spectra along the
vertical (82 for SparsePak, 382 for PMAS). The location of each spectrum
along the vertical axis is determined by finding the peak intensity
in a sliding window of fixed size oriented vertically and positioned
at the centre of the dispersion axis. Once the peaks in the centre
column are identified, we next use them as a starting point for finding
all of the other spectral pixels for each of the fibre spectra by
recursively applying the previous procedure to all of the pixels along
the dispersion axis of the spectrograph. The fibre spectra are then
extracted using the locations of the peak intensities and compiled
into a row-stacked spectrum (RSS) such that the x-axis corresponds
to the dispersion axis of the spectrograph, and the rows along the
y-axis are the fibre spectra listed in IFU order.

\paragraph{Removing Distortions and Wavelength Calibration}

The parabolic deflection caused by the inhomogeneous dispersion of
light from the pseudoslit of the IFU fibre bundle as well as fibre-to-fiber
distortions which are strongest on the edges of the detector are removed
using a two-step process. First, an emission line is chosen from the
arc lamp exposure which has strong brightness across all the fibres
(i.e., doesn't suffer from vignetting). The location of this emission
line is then traced in each fibre spectrum of the calibration exposure,
and the spectra are linearly shifted to a common dispersion axis such
that the selected emission line is located at the same wavelength
in each spectrum. A second-order correction using a fourth-order polynomial
fit is then used to recentre all of the other emission features. The
amount by which each spectral pixel is shifted along the dispersion
axis is recorded in the dispersion solution. The observed fibre spectra
are then shifted and re-entered in accord with the distortion solution
found using the calibration exposure. A standard wavelength calibration
is then applied.

\paragraph{Flat-fielding, Sky Subtraction, and Flux Calibration}

\label{sub:skySubtraction}A pseudospectrum is constructed for each
observed fibre spectrum consisting of the median-combined flux at
each wavelength from all of the fibres in the continuum exposure.
The fibre spectra are then fibre-flattened by dividing each by this
pseudospectrum to remove fibre-to-fiber transmission, unevenness in
the projection of the fibres through the pseudoslit, and the inhomogeneous
dispersion of light from the pseudoslit (see \citet{sanchez2006}
for a thorough discussion of these effects). The effects due to vignetting
are then corrected by masking any pixels in the fibre-flatted spectra
having less than seventy percent of the intensity at the same location
in the pseudospectrum. Next, a new pseudospectrum is constructed by
linearly interpolating the spectra from the sky fibres (36 for PMAS,
7 for SparsePak) onto a grid the same size as the IFU (382 for PMAS,
and 82 for SparsePak), and sky-subtraction is performed on each observed
fibre spectrum (331 for PMAS, 75 for SparsePak) using these pseudospectra.
Finally, the observed fibre spectra are flux-calibrated using spectrophotometric
standard stars acquired during each object's observing run.

\paragraph{Constructing Data Cubes}

\begin{figure*}
\begin{tabular}{cc}
\includegraphics[scale=0.95]{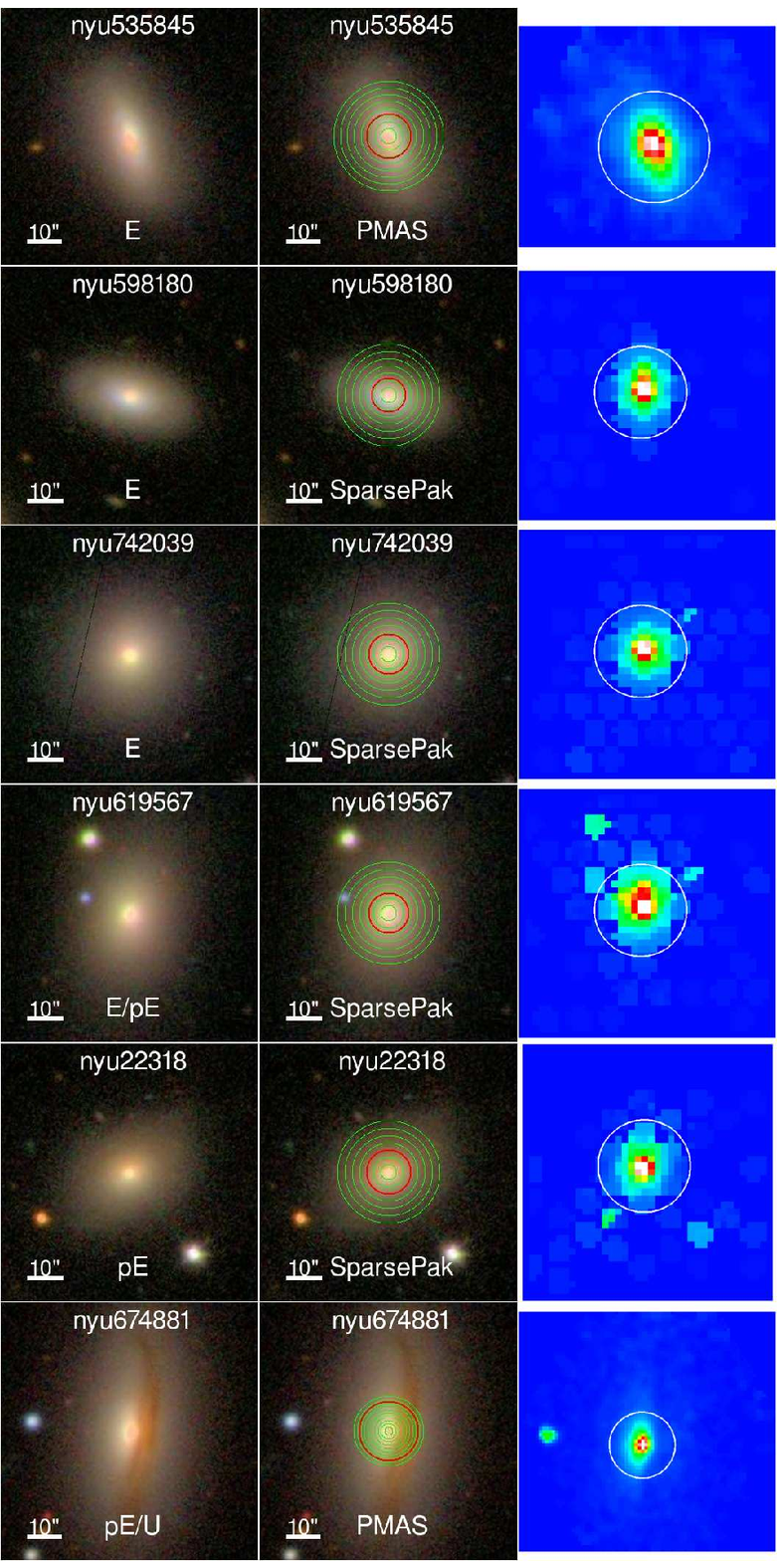} & \includegraphics[scale=0.95]{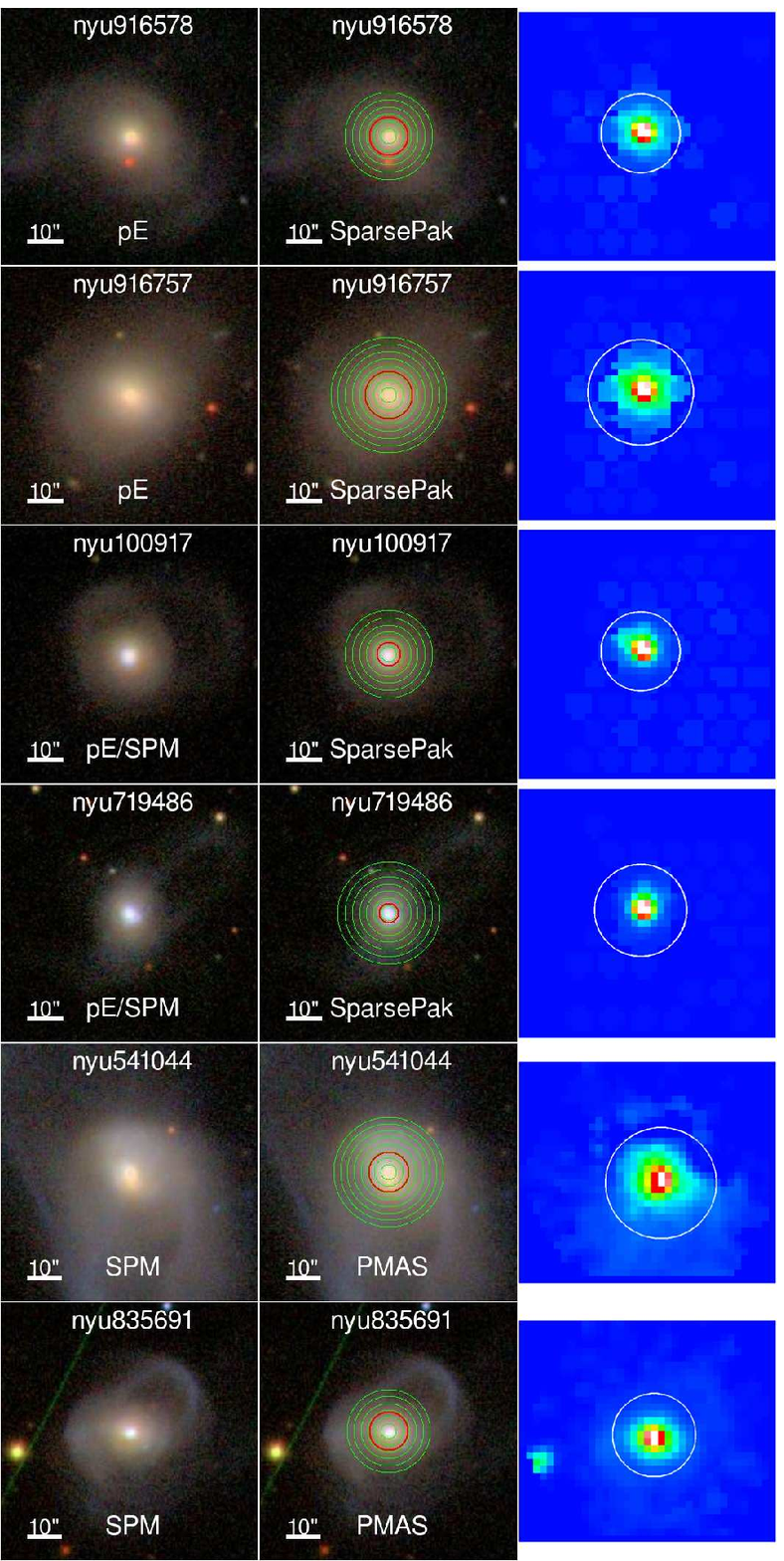}\tabularnewline
\end{tabular}

\protect\caption{\label{fig:masterMosaic}Images of our sample of galaxies. First column:
the SDSS thumbnail images scaled to the field of view of the respective
spectrograph from which the data were obtained with the morphological
classification from M14 shown bottom centre. The white bar indicates
10'' on the sky. Second column: the image from the first column with
the binned spectral annuli overlaid and the spectrograph labeled.
The red circle represents the half-light radius, $\text{R}_{50}$.
Third column: intensity in each spaxel of the data cube at rest frame
4050\AA. The white circle marks the largest annulus seen in the second
column.}
\end{figure*}

\label{sub:constructingDataCube}The wavelength- and flux-calibrated
spectra are converted into a data cube by first populating the pixel
space of the first two dimensions of the data cube with discretely-sampled
points via a standard astrometric transformation with the resulting
spaxels (SPAtial piXELS) mapping the fibre spectra onto the sky. Second,
a smooth, spatially-resolved spectroscopic light profile of each spaxel
plane is constructed for each wavelength by blending the light of
the discrete points via a two-dimensional Gaussian with a constant
circular aperture (see column three of Fig. \ref{fig:masterMosaic}).
Finally, galactic extinction along the line of sight to each target
is corrected using the Charlot and Fall \citep{charlot2000asimple}
law combining the g-band extinction values given in the SDSS with
Table 22 from \citet{stoughton2002} and Equation B1 from \citet{schlegel1998}.

\paragraph{Removing Velocity Shifts}

\label{sub:losv}In this work, we examine large extents of each galaxy's
surface where net rotation can contribute substantially to the Doppler
shift. Disentangling the component of redshift due to rotation and
recessional velocity is a complex task that we do not attempt here.
Instead, we fit a combination of one early-type and one late-type
spectral energy distribution (SED) from \citet{vazdekis2010evolutionary}
to the spectrum at each spaxel in the data cube using the FIT3D code
\citep{sanchez2011} to generate a velocity map containing spatially-resolved
redshift information over the surface of the galaxy. We input into
FIT3D a starting value of the redshift based on a spectroscopic redshift
from either SDSS or NED ($z_{published}$ from Table \ref{tab:sample}),
and the possible range of allowed redshifts given by assumption of
a maximum rotation speed of $300\;{\rm km\; s^{-1}}$ such that our
measured redshifts are in the range $z_{published}\pm300\; km\; s^{-1}/c$.
We then use the resulting velocity map to remove the Doppler shift
from the spectra. We find excellent agreement between our measured
redshifts and those published in the SDSS or NED with a maximum absolute
difference of $\sim30\;{\rm km\; s^{-1}}$. The spectra in each data
cube are then shifted to a common rest frame such that they all begin
at the same wavelength to ensure that co-adding spectra works correctly
later.

\subsubsection{Constructing Azimuthally-binned Spectra}

\label{sub:constructingBinnedSpectra}In this work, we examine the
stellar populations over large extents of each galaxy in our sample.
There are several methods available for combining IFU spectra to extract
spatially-resolved stellar populations. For example, the CALIFA collaboration
use a Voronoi tessellation \citep{cappellari2003adaptive} method
to construct local regions with high signal-to-noise \citep{cidfernandes2013resolving}.
We utilize a light-weighted azimuthal binning with constant annular
width to provide a uniform view of the radial behaviour and concurrently
counteract the decreasing signal-to-noise at large radii due to lower
surface brightness. This method has been recently used to extract
detailed Lyman alpha properties using SSP models \citep{papaderos2013nebular}
and has been shown to have a nominal effect on the derived stellar
population properties \citep{mast2014theeffects}. We use a two-stage
process for measuring the S/N/pixel of our spectra. First, we use
the standard method based on counting statistics to ensure only spectra
with S/N above the photon-limited regime are considered. Importantly,
we measure the S/N/pixel near the Balmer and 4000 break indices used
throughout this work. Second, we use the technique from \citet{husemann2013}
to account for the cross-correlated noise introduced by co-adding
spectra. In this step, we measure the S/N/pixel in several continuum
regions to ensure that our spectra are of generally high quality.

\paragraph{Error Estimate Using Counting Statistics}

Starting with the azimuthally-binned spectra, we utilize the signal
(in counts) and the noise (in counts) at each wavelength to calculate
the S/N/pixel. To do this, we first construct a row-stacked spectrum
(RSS) of the estimated Poisson noise via the standard formula $noise=\sqrt{source+background+readout^{2}}$
where the source and background are the raw (i.e., not flux-calibrated)
spectrum and sky counts, respectively, and \textit{readout} is the
readout noise from the detector. We then convert the raw spectra and
the noise RSS into data cubes following the procedure outlined in
Section \ref{sub:constructingDataCube} and apply the velocity shift
corrections discussed previously. Finally, for each annulus we construct
the azimuthal sum of the spaxels in the raw spectra data cube and
the quadrature sum of the corresponding spaxels from the noise data
cube to compute the S/N/pixel in the wavelength regions around the
${\rm D_{n}(4000)}$, H$\gamma$, and H$\delta$ indices. Because
absorption features have intrinsically less S/N/pixel than equivalent
continuum measurements, we retain annular spectra with a minimum S/N/pixel
of 5 in each region around the three indices.

\paragraph{Full Spectral Fitting}

Our analysis is centres on the equivalent width measurements of the
Balmer H$\gamma$ and H$\delta$ indices. To measure these accurately,
we must first determine the amount of nebular contamination present
in our spectra. To measure this, we use the full-spectral fitting
code FIT3D \citep{sanchez2011} and a catalogue of thirty SEDs to
find a combination which reproduces our spectra in the minimum chi-squared
sense. We use the Single Stellar Population (SSP) SEDs generated from
the MILES library using the code from \citet{vazdekis2010evolutionary}
with metallicities of 2, 20, and 100 percent solar each spanning a
range of ten ages from $\sim70\unit{\; Myrs}$ to $\sim14\;\unit{Gyrs}$.
We first fit our spectra with all of the Balmer and forbidden oxygen
emission lines masked. The total velocity dispersion (i.e., stellar
and instrumental) and internal dust extinction%
\footnote{Foreground extinction has already been removed at this point.%
} are treated as free parameters and fit in the minimum chi-squared
sense. The generated continuum fits are then subtracted from the spectra,
and the emission lines are fit.

We fit a single Gaussian to each emission line individually with the
line centre, full width half maximum, and peak amplitude treated as
free parameters. We attempt to fit all of the Balmer and forbidden
oxygen lines masked during the continuum fitting, but require that
the peak of each emission fit be greater than one sigma of the flux
contained in a window of 20 around the known line centre to prevent
fitting noise. We reject residual sky lines and any remaining cosmic
rays using a windowed boxcar filter to replace the flux exceeding
3 sigma of the flux in the window with the average of the flux in
the window. The emission fits are then subtracted from the original
spectra, and the FIT3D continuum fits are performed again with none
of the emission lines masked. This new continuum fit is then used
to perform the emission line fitting and clipping again to produce
a set of emission-free spectra.

\paragraph{Constructing Variance Maps}

During the binning process spatially coherent, but physically independent
spectra are co-added to increase the S/N per pixel. By combining independent
spectra, an artificial noise is introduced into the final spectrum
we wish to use to make our measureme\textcolor{black}{nts. To trace
this spatially-correlated noise, we use the methods of \citet{husemann2013}
to construct a }\textit{\textcolor{black}{variance map}}\textcolor{black}{{}
to estimate the correlated noise at each wavelength using the SED
fits to our emission-subtracted spectra calculated above. The variance
map is simply the median-smoothed residuals from the full spectral
fitting. Under the assumption that our statistical method provides
the ``true'' measure of the noise in the spectra, the variance map
is considered to be a direct measure of the correlated noise introduced
during the azimuthal binning. We then use the variance map as the
final measure of the error to recompute the S/N/pixel. For this final
step, we use the three regions (4040,4060\AA, 415}0,4200\AA, 4380,4430\AA)
which measure the S/N/pixel in the continuum (rather than in the indices)
where there is less change in model-data discrepancies to improve
our index measurements. Because we are measuring only continuum regions
here, we require a greater minimum S/N/pixel than was considered when
constructing error estimates using counting statistics where absorption
features were used to measure the S/N/pixel. Only spectra with S/N/pixel
> 10 are retained for further analysis. We show our sample with the
annuli overlaid on the SDSS images and the intensity map from the
data cube at 4050\AA\,  in Fig. \ref{fig:masterMosaic} and the final
set of annular spectra for each galaxy in Fig. \ref{fig:emSubSpectra}.

\subsubsection{Index Measurements}

\label{sub:measuringIndices}
\begin{figure}
\begin{centering}
\includegraphics{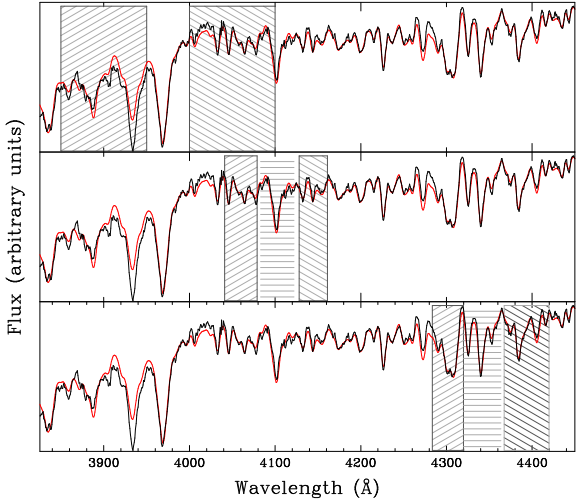}
\par\end{centering}

\protect\caption{\label{fig:lickIndices}Definitions of the blue continuum (upwards-hatched),
central (horizontal-hatched), and red continuum (downwards-hatched)
passbands for the ${\rm D_{n}(4000)}$ (top), ${\rm H}\delta_{{\rm A}}$
(centre), and ${\rm H}\gamma_{{\rm A}}$ (bottom) indices from an
example galaxy. The measured spectrum is shown in black, and the continuum
fit is shown in red. Note that the ${\rm D_{n}(4000)}$ index definition
does not include a central continuum. The y-axis shows the flux arbitrarily
scaled to show detail, and the x-axis shows the rest frame wavelengths.}
\end{figure}

Starting with the set of fully reduced
and calibrated emission-subtracted annular spectra in Appendix \ref{sec:app_radialSpectra},
we measure the ${\rm H}\delta_{{\rm A}}$ and ${\rm H}\gamma_{{\rm A}}$
Lick indices and the narrow-band 4000\AA\ break as a function of
galactic radius. To calculate the strength of each index, we first
define three passbands around each index's line centre: the blue continuum,
the central region, and the red continuum. The passbands for the Lick
${\rm H}\delta_{{\rm A}}$ and ${\rm H}\gamma_{{\rm A}}$ indices
from \citet{worthey1997hgamma} and the narrow-band 4000\AA\ break
(hereafter referred to as ${\rm D_{n}(4000)}$) from \citet{balogh1999}
are shown in Fig. \ref{fig:lickIndices}. To reduce random errors
in the outermost annuli of our galaxies, we use the average Balmer
index $\left\langle {\rm H}\delta_{{\rm A}},{\rm H}\gamma_{{\rm A}}\right\rangle \equiv\left({\rm {\rm H\delta_{{\rm A}}+{\rm H\gamma_{{\rm A}}}}}\right)/2$
from \citet{sanchez-blazquez2006stellar}.

\begin{figure*}
\includegraphics{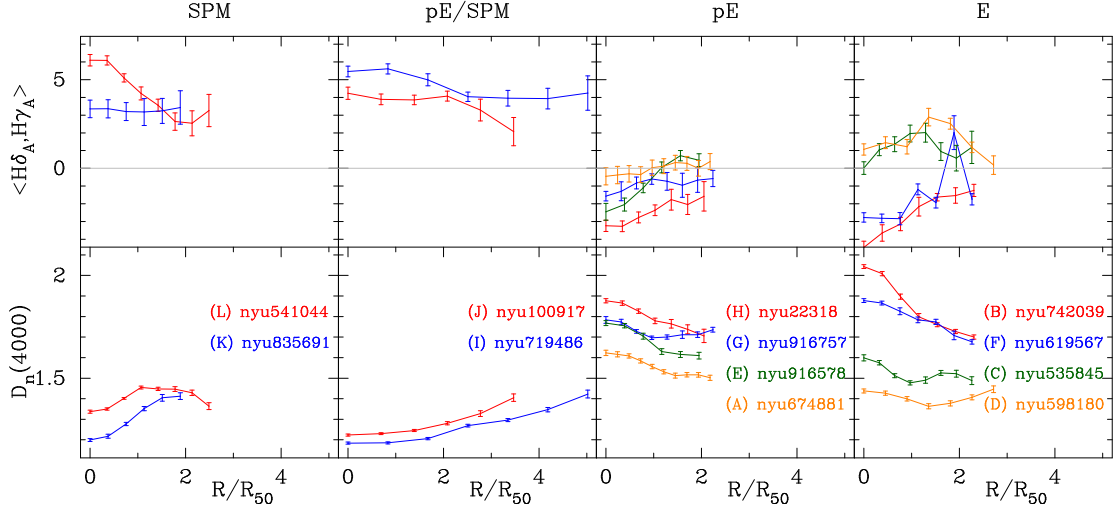}\protect\caption{\label{fig:indicesByMorphType}Average stellar Balmer absorption indices,
$\left\langle {\rm H}\delta_{{\rm A}},{\rm H}\gamma_{{\rm A}}\right\rangle \equiv\left({\rm {\rm H\delta_{{\rm A}}+{\rm H\gamma_{{\rm A}}}}}\right)/2$,
and the 4000\AA\  break strength, ${\rm D_{n}(4000)}$, as a function
of the SDSS Petrosian r-band half-light radius (R\protect\textsubscript{50})
organized by morphological classification. Colours represent indices
from the same galaxy in a single column.}
\end{figure*}

In general, differences in ${\rm D_{n}(4000)}$ correspond to differences
in the ages of stellar populations on the order of several Gyrs, and
differences in Balmer strength correspond to differing fractions of
A stars. We show our index measurements as a function of galactic
radius organized by morphological classification in Fig. \ref{fig:indicesByMorphType}.
Examining the radial trends of the indices, we find that strongly
disturbed galaxies (visually classified as SPM or pE/SPM) exhibit
flat or decreasing Balmer absorption at increasing radii with generally
increasing ${\rm D_{n}(4000)}$ values with increasing galactic radius--
indicating that the youngest stellar populations are in the galaxies'
cores. For the modestly disturbed galaxies (pE) and two of the smooth
ellipticals (E) nyu742039 and nyu619567, we find generally decreasing
${\rm D_{n}(4000)}$ and increasing Balmer values at increasing radii
consistent with their oldest stellar populations residing in their
cores. We note that most of the Balmer radial indices are negative
and hence not indicative of stellar populations containing a large
fraction of A stars. However, there are two exceptions: nyu674881
and nyu916578 have positive Balmer absorption values at their largest
radii. The final two smooth ellipticals (E) nyu598180 and nyu535845
are distinct in that they have positive Balmer absorption values at
nearly all radii and small ${\rm D_{n}(4000)}$ values near the range
of values exhibited by the SPM galaxies. From Fig. \ref{fig:masterMosaic},
we see that these two galaxies possess blue circumnuclear rings consistent
with recent star formation. As discussed in Section \ref{sub:sampleSelection},
our choice to include galaxies with interesting morphological details
such as the blue rings seen here has paid dividends in that we find
there need not be a direct correlation between a galaxy's general
morphology and its \textit{recent} star formation history.

\section{Modelling Star Formation Histories}

\label{sec:modellingSFHs}In this work, we seek to test if the modern
merger hypothesis can account for the assembly of our sample of plausible
new ellipticals with unusually blue optical colours. Specifically,
we use the prediction that the blue-to-red migration leaves behind
a tell-tale signature via the presence of a strong, centrally-concentrated
burst of star formation in the merger remnant. This signature can
be detected in our sample of plausible new ellipticals by comparing
the measured Balmer and ${\rm D_{n}\left(4000\right)}$ indices (see
Section \ref{sub:measuringIndices}) to a suite of specially-constructed
models.

We use the iSEDFit tool from the \textsc{impro} code package%
\footnote{https://github.com/moustakas/impro%
} to construct a suite of theoretical spectral energy distributions
(SEDs) covering many different SFHs \citep[see appendix A of][for details]{moustakas2013primusconstraints}.
We begin construction by characterizing the star formation rate (SFR)
as a function of time for an arbitrary stellar population using a
simple, exponentially-declining SFR such that ${\rm SFR}\left(t\right)\propto e^{-t/\tau}$
where $\tau$ is the characteristic time. We then allow for two possible
evolution paths: one in which the galaxy continuously forms stars
throughout its lifetime and a second in which the galaxy experiences
a burst of star formation superimposed atop a continuously-declining
SFH (see Appendix \ref{sec:app_constructingBurstModels} for details).

The continuous models begin with a large burst of star formation at
$t=0$ and evolve for $\sim15\;{\rm Gyrs}$ with a continuous, exponentially-declining
SFR. For these models, we allow a range of e-folding times from $7{\rm \; Gyrs}$
representative of normal star-forming late-type galaxies to $0.5\;{\rm Gyrs}$
signifying a quenching of star formation. The burst models start with
the same large burst at $t=0$ as in the continuous models but add
an additional burst of star formation generating between $1\%$ and
$25\%$ of the total mass of the galaxy atop the exponentially-declining
star formation when the galaxy is $\sim12\;{\rm Gyrs}$ old. We use
e-folding times from $25\,{\rm Myrs}$ to $1{\rm \, Gyr}$ representative
of strongly- and weakly-truncated bursts of star formation, respectively.
These burst models reflect the scenario in which our plausible new
ellipticals formed from a \textit{recent} (${\rm z=0.12}$) gas-rich
merger and evolved to today with varying degrees of quenching over
a period of evolution on the order of the dynamical merger time ($\sim1\;{\rm Gyr}$;
\citealt{lotz2008}). To produce the final model spectra, we convolve
the constructed SFHs with a set of single stellar population SEDs
from \citet{bruzual2003} spanning their full range of metallicities
from $0.005\,{\rm Z_{\odot}}$ to $2.5\,{\rm Z_{\odot}}$ with the
Chabrier IMF.

\subsection{Index Plane}

\label{sub:indexPlane}
\begin{figure*}
\begin{centering}
\includegraphics{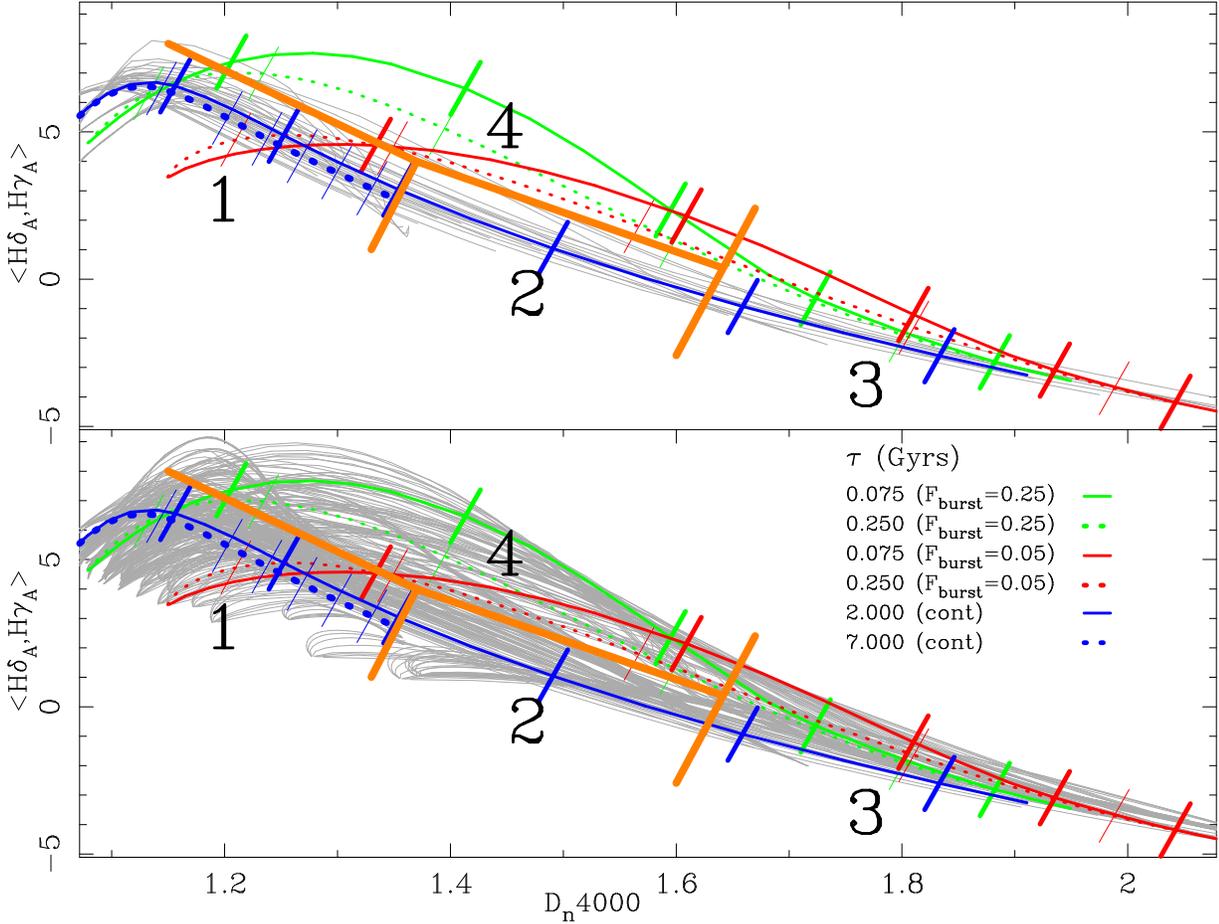}
\par\end{centering}

\protect\caption{\label{fig:splitModelPlane}The stellar Balmer absorption versus 4000\AA\
break strength model plane showing two continuous SFH models (blue),
two burst models in which 25\% of the total galaxy mass was formed
in a single burst (green), and two burst models with a 5\% burst fraction
(red). Tick marks denote ages along the model lines. For the continuous
models, the ages are 1, 3, 5, 7, 9, and 11 Gyrs. For the burst models,
the ages (i.e., time since the burst) are 0.2, 0.5, 0.9, 1.5, and
2.5 Gyrs. The four qualitative regions from the text are labelled
and orange lines mark their approximate boundaries. The full suite
of continuous (top) and burst models (bottom) covering their respective
full parameter spaces are shown in light grey.}
\end{figure*}

We use the established technique of comparing
the 4000\AA\  break strength with the Balmer Lick indices ${\rm H\gamma_{A}}$
and ${\rm H\delta_{A}}$ to qualitatively distinguish between galaxies
that have a simple, continuous SFH from those that have undergone
a burst of star formation within the last $\sim1\;{\rm Gyr}$ \citep[K03]{kauffmann2003}.
Figure \ref{fig:splitModelPlane} shows the indices measured from
our continuous models (blue) and our burst models (green and red)
in the Balmer-${\rm D_{n}(4000)}$ plane. We note that though there
are parts of parameter space that can clearly only be reached by models
with bursts, the burst models substantially overlap the regions of
parameter space covered by the suite of continuous models; a result
of the age-burst degeneracy. The locus of points extending along any
continuous model track can be explained by a range of continuous SFHs
of different e-folding times and luminosity-weighted ages at fixed
metallicity. As shown by K03, this locus is not as sensitive to metallicity
as the indices themselves and acts as a simple diagnostic allowing
us to qualitatively separate galaxies undergoing continuous star formation
from those which have undergone a burst of star formation in the last
$1-2\;{\rm Gyrs}$ in which at least $5\%$ of their mass was formed.

To simplify our analysis, we choose our fiducial burst and continuous
models (blue, red, and green model tracks in Fig. \ref{fig:splitModelPlane})
to have solar metallicity. We choose solar metallicity for three reasons:
1) interpretation of the age and chemical composition of composite
stellar populations determined with line-strength indices depends
upon the chemical composition of the underlying 'old' stellar populations
\citep{serra2007onthe}, 2) SDSS galaxies in our mass range have \textit{unresolved}
metallicities near solar \citep{gallazzi2005}, and 3) local elliptical
galaxies have shallow gradients in their SSP-equivalent chemical compositions
\citep{trager20002}. We note, however, that within both the continuous
and burst SFH models, metallicity gradients can plausibly mimic age
gradients, and caution against overinterpreting the model plane.

Points in the plane which lie above the continuous models (i.e., have
a more positive Balmer absorption index at a fixed ${\rm D_{n}(4000)}$)
can be explained by a range of burst SFHs of different e-folding times,
luminosity-weighted ages at fixed metallicity, and fraction of new
stars formed during the added burst. Because of this degeneracy we
do not try to precisely constrain model parameters such as e-folding
times, etc. Yet, by comparing the data to our chosen models sampling
the extrema of reasonable e-folding times and luminosity-weighted
ages, we can draw several qualitative conclusions. To do this, we
divide the Balmer-${\rm D_{n}(4000)}$ parameter space into four regions
by combining the known properties of the constructed models with the
regions of the Balmer-${\rm D_{n}(4000)}$ plane where K03 found $95\%$
of their simulated galaxies had high probabilities of forming either
at least $5\%$ or exactly none of their stars in a burst within the
last $2\unit{{\rm Gyrs}}$ (see figure 6 in K03 for details).

The first region (\textbf{Region 1}) in the model plane contains the
radial indices with ${\rm {\rm D_{n}(4000)}}\lesssim1.35$ that lie
on or below the continuous SFH models where typical late-type star-forming
galaxies are found. We note that points lying along the burst models
having ages $\lesssim200\;{\rm Myrs}$ also reside in this region.
This is an example of the well-known age-burst degeneracy wherein
a large burst of star formation occurring long ago (here, the continuous
model) mimics the behaviour of a recent, smaller burst. The second
region (\textbf{Region 2}) contains the range ${\rm 1.35\lesssim{\rm D_{n}(4000)}}\lesssim1.6$
and lies on or below the continuous SFH model with a short e-folding
time ($\tau=2\;{\rm Gyrs}$, solid blue lines). In this region, we
find galaxies having luminosity-weighted ages between $\sim5$ and
$9\;{\rm Gyrs}$, but note that adding a few percent of new stars
by mass atop an older population can mimic these ages.

The third region (\textbf{Region 3}) contains the radial indices with
${\rm D_{n}(4000)}\gtrsim1.6$ and lying in the region where the continuous
and burst models overlap. At fixed ${\rm D_{n}(4000)}$, measurements
in this region have Balmer absorption indices which span from the
continuous SFHs consistent with old stellar populations in quiescent
ETGs to the $5\%$ burst models with luminosity-weighted ages $\apprge1\,{\rm Gyrs}$.
Finally, the fourth region (\textbf{Region 4}) spans ${\rm D_{n}(4000)}\lesssim1.6$
and is bracketed above the continuous SFH models. This region is described
by K03 as containing galaxies with a high confidence of having undergone
a burst of star formation in the last $\sim2\unit{{\rm \, Gyrs}}$.
The individual Balmer-${\rm D_{n}(4000)}$ planes with the radial
indices for each galaxy in our sample are shown in Fig. \ref{fig:modelComparison}.

\subsection{Comparing Models and Data}

\begin{figure*}
\centering{}\includegraphics{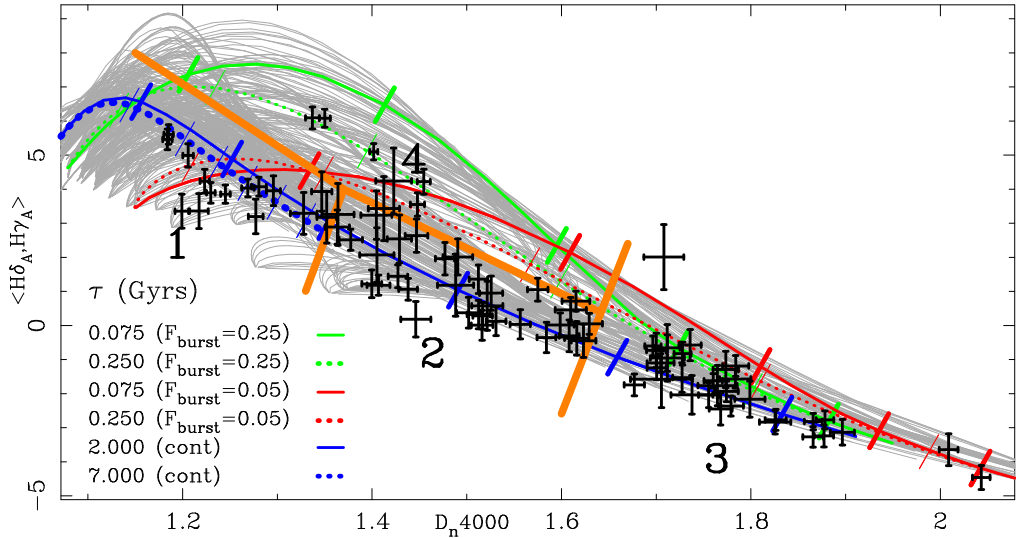}\protect\caption{\label{fig:modelPlaneWithPoints}The stellar Balmer absorption versus
4000\AA\ break strength model plane as from Fig. \ref{fig:splitModelPlane}
with our measured radial indices and their errors for all radii (see
Fig. \ref{fig:modelComparison}) of our entire sample are shown in
black. The complete suite of continuous and burst models are shown
in grey. The four qualitative regions from the text are labelled and
orange lines mark their approximate boundaries.}
\end{figure*}

In Fig. \ref{fig:modelPlaneWithPoints}, we overlay all of our radial
index measurements from all galaxies in our sample atop our full suite
of models and find two notable conclusions. First, our measured indices
span nearly the entire model space indicating that our sample of galaxies
possesses a wide range of SFHs (although radial information is not
recoverable here). Second, our chosen model space completely covers
our measured indices%
\footnote{The one outlier near ${\rm D_{n}(4000)\simeq1.7}$ is shown in Section
\ref{sub:formationHistory_manyPaths} to likely be a measurement error.%
}. This indicates that our choice of model parameters is congruent
with the local population of plausible new ellipticals. We note that
many of the points that lie in the fourth region of the plane (i.e.,
bracketed above the continuous models (blue)) reside below the burst
models with burst fractions of $5\%$ (red) and are more consistent
with burst models with smaller burst fractions (e.g., $\sim1\%$).
We defer to future work to recover detailed star formation histories
for individual data points.

\begin{figure*}
\centering{}\includegraphics{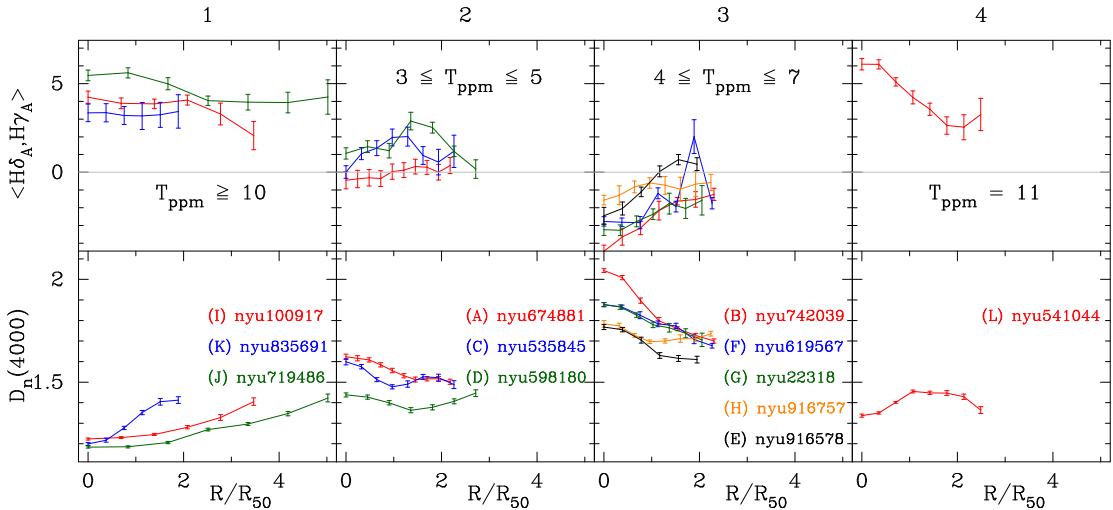}\protect\caption{\label{fig:indicesByRegion}The average stellar Balmer absorption
indices, $\left\langle {\rm H}\delta_{{\rm A}},{\rm H}\gamma_{{\rm A}}\right\rangle \equiv\left({\rm {\rm H\delta_{{\rm A}}+{\rm H\gamma_{{\rm A}}}}}\right)/2$,
and the 4000\AA\  break strength, ${\rm D_{n}(4000)}$, at each annulus
shown as a fraction of the SDSS Petrosian r-band half-light radius
(R\protect\textsubscript{50}) for each galaxy organized by location
in the Balmer-${\rm D_{n}4000}$ plane (see Appendix \ref{sec:app_indexModelPlane}).
Colours represent indices from the same galaxy in a single column.
The range of ${\rm T_{ppm}}$ values from Table \ref{tab:sample}
are shown for the galaxies contained in each region of the plane where
larger values indicate stronger morphological disturbances.}
\end{figure*}

In Fig. \ref{fig:indicesByRegion}, we show the Balmer and ${\rm D_{n}(4000)}$
radial indices for each of our galaxies organized by the region in
which most of the indices lie (see Appendix \ref{sec:app_indexModelPlane}
for individual Balmer/${\rm D_{n}\left(4000\right)}$ planes). Galaxies
with radial indices in two different regions of the plane (e.g., nyu916578)
are assigned to the region where their central annuli lie. Comparing
with Fig. \ref{fig:indicesByMorphType}, we see that the first and
fourth regions of the Balmer-${\rm D_{n}(4000)}$ plane contain the
most morphologically disturbed galaxies (${\rm T_{ppm}}\ge10$) with
flat or slightly decreasing Balmer index gradients and generally increasing
${\rm D_{n}(4000)}$ indices at large radii. With burst-like SFHs
and very young stellar ages in their cores, these galaxies are likely
candidates of recent gas-rich major mergers. If the dynamical disturbance
is due to merging activity as predicted by the modern merger hypothesis,
the dichotomy of regions in the plane where strongly disturbed galaxies
lie indicates that not all mergers result in starbursts- consistent
with predictions from simulations \citep[e.g., ][]{cox2008}. Galaxies
in region 2 have smooth elliptical morphologies (${\rm T_{ppm}\leq5)}$
and SFHs consistent with weak, but ongoing, star formation. Galaxies
in region 3 have steep positive Balmer index gradients and steep negative
${\rm D_{n}(4000)}$ gradients indicative of very old stellar ages
in their central regions, implying they are the least likely to be
remnants of recent gas-rich major mergers. Surprisingly, these galaxies
exhibit a variety of elliptical morphologies (${\rm 4\le T_{ppm}\le7}$)
including some with clear signs of interaction (e.g., nyu916578),
but possess stellar ages much older than those found in the smooth
ellipticals in region 2.

\section{Dissecting Assembly Histories}

\label{sec:dissectingFormationHistories}A key prediction of the modern
merger hypothesis is the presence of a strong, centrally-concentrated
burst of star formation in the merger remnant. We anticipate this
inside-out growth produces a distinct radial SFH which we can use
to qualitatively differentiate plausible remnants of major mergers
from galaxies undergoing other mass assembly mechanisms. Combining
our spectroscopic radial index measurements with our qualitatively
defined regions of the Balmer-${\rm D_{n}(4000)}$ plane (see Section
\ref{sub:indexPlane}), we ask the question whether each galaxy in
our sample is consistent with the predictions of the modern merger
hypothesis; i.e., having preferentially younger stars in its core
formed in a recent burst.

Additionally, we anticipate central star formation to alter the galaxy's
optical colours such that it either preferentially contains more blue
light in its core or its core colours are reddened consistent with
the presence of merger-induced dust. We therefore examine the distribution
of global colours using the SDSS (\textit{u-r}) and (\textit{r-z})
model magnitudes (Fig. \ref{fig:colorColor}, left panel) and core
colours using the SDSS fibre magnitudes (right panel); both are k-corrected
to redshift zero in the AB magnitude system. For comparison, we show
the control samples from M14 of\textbf{ }red, spectroscopically quiescent
ETGs and blue, star-forming LTGs from the SDSS with the same mass
and redshift range as our sample. For the core colours, we show only
those galaxies with ${\rm z}\le0.03$ to match the redshift range
of our sample and limit the spatial bias of the SDSS fibre aperture%
\footnote{For the redshift range of our targets, the 3\arcsec\ SDSS fibre aperture
corresponds to between 0.86 and 1.8 kpc.%
}. In both panels, we show the empirical boundary from \citet{holden2012}
which robustly separates star-forming red (dusty) galaxies from non
star-forming (old) galaxies. M14 show that $92\%$ of all spectroscopically
quiescent galaxies from their complete sample lie above this line
and $97\%$ of BPT star-formers lie below it. Additionally, we overlay
the sample of spectroscopically-confirmed star-forming ellipticals
from the M14 sample as blue asterisks.

In Fig. \ref{fig:colorGradients}, we provide quantitative colour
gradients in the SDSS (\textit{u-r}) and (\textit{r-z}) colour-colour
plane for our galaxies and the subset of the comparison samples shown
in Fig. \ref{fig:colorColor} with $z\le0.03$. The colour gradients
are defined by $\Delta C=C{}_{{\rm outer}}-C_{{\rm core}}$ where
$C$ is one of (\textit{u-r}) or (\textit{r-z}). The outer colour
$C_{{\rm outer}}$ is derived by converting the respective SDSS model
magnitudes into flux, subtracting the flux from the SDSS fibre magnitudes,
and converting to a colour. We note that the colour gradients computed
here are not the same as the tracks shown in the right-hand panel
of Fig. \ref{fig:colorColor} which show the relative position of
each galaxy in the colour-colour plane according to its core and global
colours.

\subsection{Are starbursts always concurrent with major mergers?}

\label{sub:formationHistory_starburst}
\begin{figure*}
\begin{centering}
\includegraphics{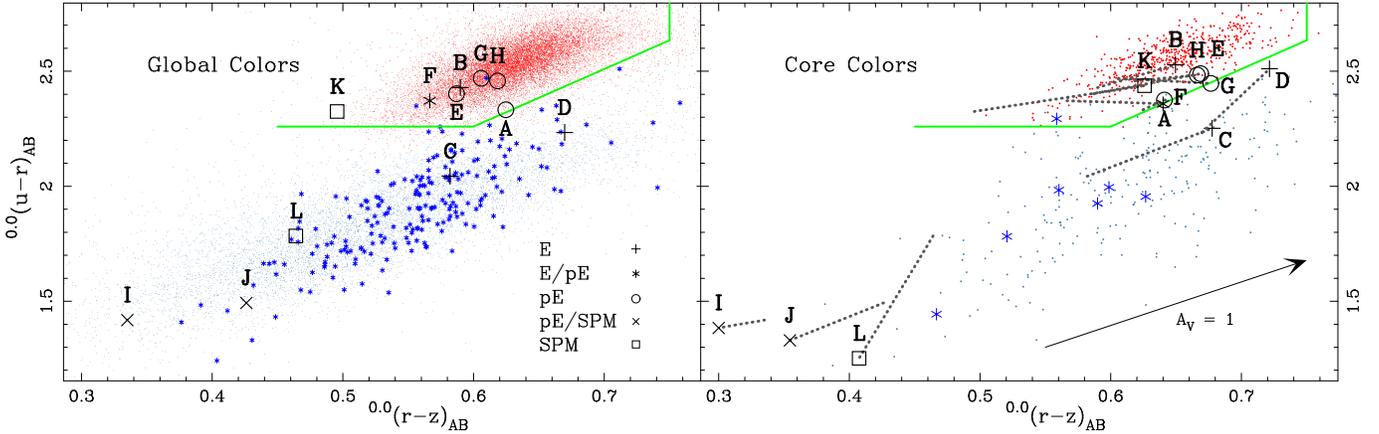}
\par\end{centering}

\protect\caption{\label{fig:colorColor}{[}left{]} Global (\textit{u-r}) and (\textit{r-z})
colours and {[}right{]} core colours from the SDSS k-corrected to
redshift zero in the AB magnitude system over-plotted on the quiescent
(red) and star-forming (blue) control samples described in Section
\ref{sec:dissectingFormationHistories}. Core colours are shown only
for galaxies with $z\le0.03$ to match the redshift of our sample
and minimize aperture effects. The dotted lines in the figure on the
right connect the core colours with the corresponding global colours
for each galaxy. Symbols represent visual classifications from M14.
The empirical boundary from \citet{holden2012} (green) robustly separates
the star-forming and non star-forming galaxies with star-forming galaxies
below the line. The complete sample of star-forming elliptical galaxies
from M14 is shown in blue asterisks. Letter labels are from Table
\ref{tab:sample}.}
\end{figure*}

\begin{figure}
\begin{centering}
\includegraphics{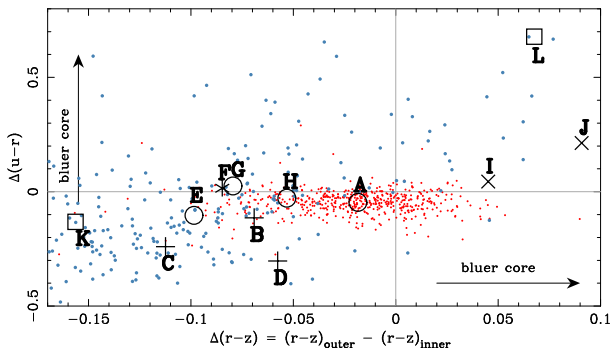}
\par\end{centering}

\protect\caption{\label{fig:colorGradients}The SDSS model (\textit{u-r}) and (\textit{r-z})
colour gradients for massive (${\rm {\rm \log M_{*}\ge10}}$), nearby
(${\rm z\le0.03}$) galaxies. We show the spectroscopically quiescent
ETG (red) and blue star-forming LTG (blue) control groups from Fig.
\ref{fig:colorColor}. Our sample of massive blue ellipticals are
shown using their respective morphological symbols and labels as in
Fig. \ref{fig:sampleSelection_colorMass}. Each 'blue core' vector
points in the direction of its respective colour gradient having a
bluer core colour.}
\end{figure}

\begin{figure*}
\begin{centering}
\includegraphics{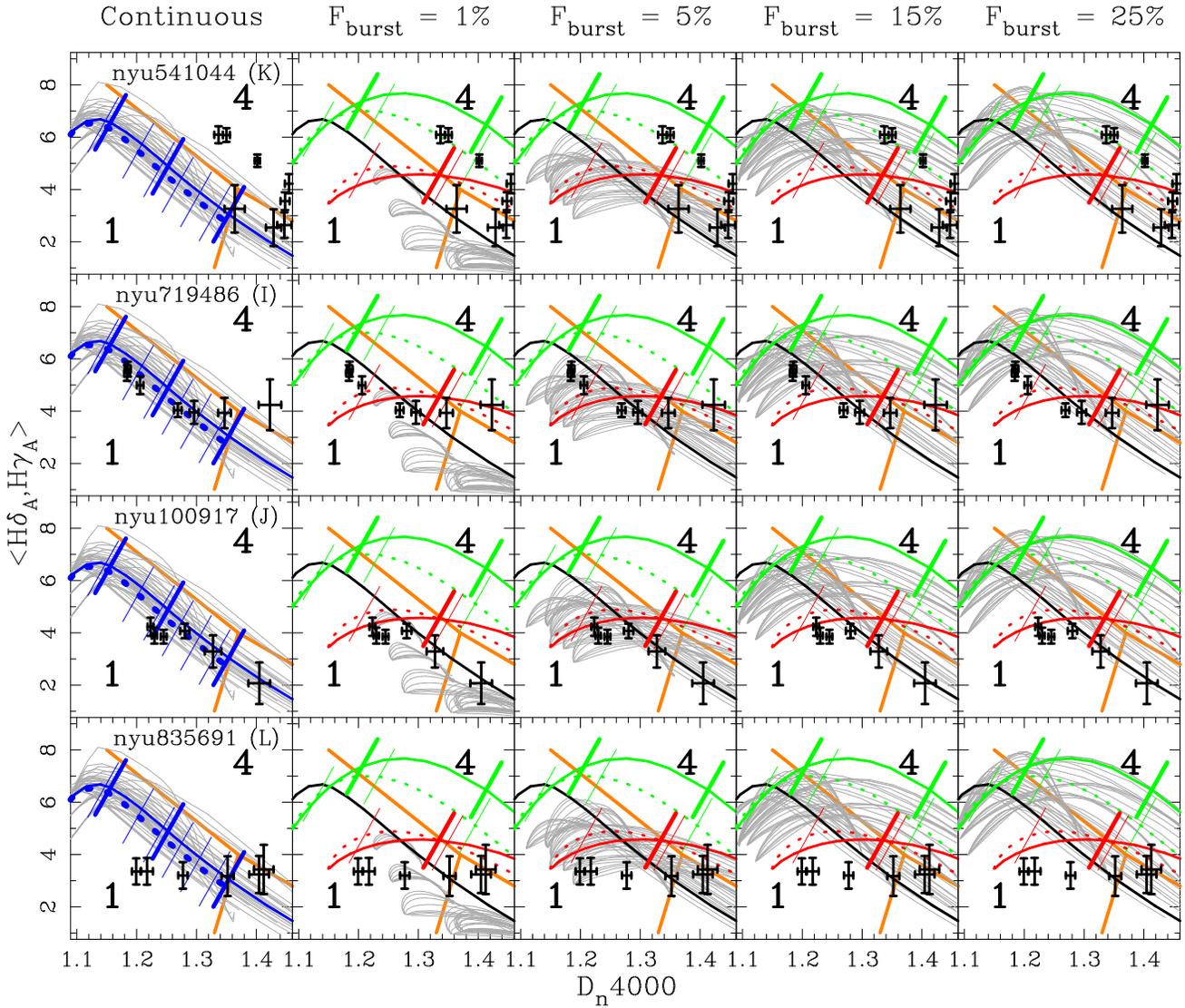}
\par\end{centering}

\protect\caption{\label{fig:region1Zoom}Zoom-in of regions 1 and 4 of the Balmer-${\rm D_{n}(4000)}$.
Column 1 shows the full suite of continuous models in grey and highlights
the two fiducial models used throughout this work in blue. Columns
2 through 5 show the full suite of burst models in grey with the fraction
of stellar mass formed during the burst increasing to the right. The
two sets of fiducial burst models used throughout this work are shown
in red ($5\%$ burst) and green ($25\%$ burst). The fiducial continuous
model with an e-folding time of $7\unit{{\rm Gyrs}}$ is shown in
solid dark grey alongside the burst models as a guide. Each row shows
the radial index measurements corresponding to the galaxy listed in
the first column.}
\end{figure*}

Examining Figs. \ref{fig:indicesByMorphType}
and \ref{fig:indicesByRegion}, we find that only one (nyu541044)
of the four galaxies visually classified as SPM or pE/SPM (i.e., ${\rm T_{ppm}\ge10}$)
has radial indices in region 4. In particular, Fig. \ref{fig:region1Zoom}
shows its central indices are consistent with our burst models in
which $\ge15\%$ of its stars formed in a recent burst and have luminosity-weighted
ages $\le1\,{\rm Gyr}$ (measured relative to the start of the burst).
Surprisingly, this galaxy's global and core optical colours both lie
in the non star-forming region of colour-colour space as determined
by the Holden boundary (Fig. \ref{fig:colorColor}). However, its
colour gradients lie outside the locus of local ETGs and are more
steeply negative (i.e., redder cores) than $\sim80\%$ of the local
star-forming LTGs (Fig. \ref{fig:colorGradients}). This combination
of clear burst signature and quiescent core colours is consistent
with post-starburst galaxies which are characterized as having recently
undergone a large burst of star formation quickly followed by strong
quenching.

Indeed, previous spectroscopic study of this galaxy by \citet{schweizer1996}
found this to be the case; although we find a slightly higher fraction
of new stars formed in the burst ($\ge15\%$ compared to their $\sim10\%$)
likely attributable to the more modern stellar template libraries
used in constructing our burst models. Galaxies with post-starburst
signatures are also referred to as E+A galaxies due to their spectra
appearing as a superposition of an elliptical galaxy (i.e., very little
nebular emission) and that of A-type stars \citep{dressler1983spectroscopy}.
From Fig. \ref{fig:emSubSpectra}, we see the presence of both an
elliptical-like spectrum as well as the Balmer absorption features
in our observed spectra for this galaxy. Additionally, the lack of
dust signatures in the optical colours of this galaxy (Fig. \ref{fig:colorColor})
is consistent with local E+A galaxies being quenched post-starbursts
rather than starbursts with dust-obscured nebular emission \citep{goto2004areea}.

The characteristics of E+A galaxies have been shown to be consistent
with the transformation from a gas-rich, star-forming disk galaxy
to a gas-poor, quiescent, pressure-supported spheroidal galaxy \citep{norton2001},
and are hypothesized to come from gas-rich major mergers \citep{wild2009,cox2008,goto2008,pracy2012}.
The molecular gas density profile of this galaxy \citep{hibbard1999luminosity}
is consistent with the centrally-concentrated gas deposition predicted
in simulations of major mergers \citep[e.g., ][]{barnes1991}. The
qualitative luminosity-weighted ages we find in its central annuli
are consistent with the lifetime of the short ($\sim0.1-0.3\,{\rm Grys}$)
E+A signatures seen in simulations \citep{snyder2011}. Without further
spectroscopic measurements, we are unable to draw any conclusions
regarding the role of an AGN as the truncation mechanism as predicted
by the modern merger hypothesis. Clearly, though, the preponderance
of evidence indicates that this galaxy is a strong candidate of being
a remnant of a recent, gas-rich major merger.

The remaining three galaxies with ${\rm T_{ppm}\ge10}$ have star-forming
global colours with bluer core colours (Fig. \ref{fig:colorColor})
and steeper positive (i.e., bluer core) colour gradients than $\sim96\%$
of the local population of star-forming LTGs (Fig. \ref{fig:colorGradients}).
These strongly disturbed galaxies have generally larger Balmer absorption
values and smaller 4000\AA\  break strengths than all of the other
morphology types (Fig. \ref{fig:indicesByMorphType}), and specifically
have their largest Balmer and smallest ${\rm D_{n}(4000)}$ values
in their centres-- consistent with young, centrally-concentrated stellar
populations. These characteristics match the core star formation coincident
with strong morphological disturbances predicted by the modern merger
hypothesis and are consistent with these galaxies forming via gas-rich
major mergers. Unlike nyu541044, however, these galaxies lack any
indication of post-starburst signatures. Rather, they lie in region
1 (Fig. \ref{fig:indicesByRegion}) of the Balmer-${\rm D_{n}(4000)}$
plane and exhibit a dichotomy of plausible SFHs from long-lived continuous
star formation to a small, recent burst of star formation. As noted
in Section \ref{sub:indexPlane}, the overlap between the continuous
and burst models in this region is a consequence of the well-known
age-burst degeneracy. The intricacies of this effect are outside the
scope of this work, yet we can constrain the plausible formation scenarios
by carefully examining region 1 over our suite of models.

In Fig. \ref{fig:region1Zoom}, we show a zoom-in of the Balmer-${\rm D_{n}\left(4000\right)}$
plane covering regions 1 and 4 with varying configurations of the
parameter space from our models. Of the three galaxies which lie in
region 1, only nyu835691 exhibits central indices which are \textit{inconsistent}
with the continuous models. For the two galaxies with central annuli
consistent with the fiducial continuous models, we note that these
models exhibit luminosity-weighted ages in region 1 ranging between
$1-11\;{\rm Gyrs}$ depending on e-folding time. However, such large
ages are in conflict with the predicted lifetimes of the tidal features
seen in these galaxies (see below). Although our observations do not
conclusively exclude the continuous models, we note that they are
only plausible for these two galaxies and do not consider them further.
Examining the burst models in columns 2 through 5, we see that all
three galaxies are generally inconsistent with the extreme models
having burst fractions of $1$ and $25\%$, respectively. For the
models with moderate burst fraction, we see that the radial indices
of all three galaxies lie along the model tracks with ${\rm F_{burst}=5\%}$
(consistent with the findings of K03 for this region), but only nyu719486
has \textit{central} indices along the model tracks with the $15\%$
burst fraction.

Focusing on the burst models with ${\rm F_{burst}=5\%}$, we find
very young luminosity-weighted ages ($\lesssim200\;{\rm Myrs}$, measured
relative to the time of the burst) over all metallicities and e-folding
times. Given that the burst models were constructed with the burst
(and thus the merger) occurring at ${\rm z=0.12}$ and our galaxies
have redshifts of ${\rm z\simeq0.03}$, this leaves an intermission
of $\sim1\;{\rm Gyr}$ between coalescence of the merger progenitors
and our observations of these galaxies (assuming our chosen model
is correct). This period is consistent with the lifetimes of the strong
tidal signatures and asymmetries seen in these galaxies as well as
in simulations of gas-rich major mergers \citep{lotz2008}.

The lack of clear signatures of a large burst of star formation in
three of the four SPM or pE/SPM (i.e., ${\rm T_{ppm}\ge10}$) galaxies
in our sample is consistent with previous studies of blue spheroids
\citep[e.g.,][]{tojeiro2013thedifferent} and actively star-forming
local ellipticals \citep{fukugita2004} which were shown to have SFHs
similar to the long-lived star formation seen in local star-forming
LTGs. This indicates that not all plausible major merger remnants
undergo an intense burst of star formation. Indeed, this dichotomy
is also seen in simulations of gas-rich major mergers which show that
the strength and duration of merger-induced starburst activity is
likely dependent upon the gas mass fraction and mass ratios of the
progenitor disks \citep{cox2008} as well as orbital parameters at
fixed masses \citep{snyder2011}. We therefore find that they are
likely candidates of being remnants of recent, gas-rich major mergers.

\subsection{Dusty ellipticals temporarily removed from the red sequence?}

\label{sub:formationHistory_dusty}As shown in Fig. \ref{fig:indicesByRegion},
galaxies with radial indices lying principally in region 2 of the
Balmer-${\rm D_{n}(4000)}$ plane exhibit relatively flat gradients
in ${\rm D_{n}(4000)}$ ($\Delta{\rm D_{n}(4000)}\simeq0.1$) but
non-uniform gradients in Balmer absorption that peak at radii $>1{\rm R_{e}}$.
The flat gradients in ${\rm D_{n}(4000)}$ with variable Balmer absorption
are indicative of a changing fraction of young ($<1\,{\rm Gyrs}$)
stars atop a uniform, older population. Such a ``frosting'' of low-level
star formation \citep{trager2000,yi2005,kaviraj2009,salim2010starformation}
may add only $1-2\%$ to a galaxy's mass, but has enough hot O and
B stars to produce a brief blue colour migrating these ellipticals
temporarily blue-ward off of the red sequence into the green valley
\citep{cortese2009,thilker2010ngc404}. Overall, these galaxies have
smooth morphologies (E), but nyu674881 is labelled pE owing to its
large, circumgalactic dust ring (Fig. \ref{fig:masterMosaic}).

Examining their colours in Fig. \ref{fig:colorColor}, we find the
two ellipticals (${\rm T_{ppm}=5}$) have global colours which lie
in the region of the colour-colour plane occupied by star-forming
LTGs as well as the spectroscopically-confirmed star-forming ellipticals
from M14. Their core colours track upward along the dust-reddening
vector away from the star-forming ellipticals toward redder colours;
suggesting the presence of central dust obscuration. Their colour
gradients (Fig. \ref{fig:colorGradients}) lie outside the locus of
quiescent ETGs with (u-r) gradients a factor of $\sim2$ steeper than
those seen in the quiescent ETGs. Such steepened colour gradients
(i.e., redder core colours) have been previously identified in nearby,
luminous elliptical galaxies with optical dust features \citep{kim2013opticalnear}
and are likely seeded from accretion of a gas-rich satellite \citep{martini2013theorigin}.
Additionally, these galaxies possess circumnuclear blue rings (Fig.
\ref{fig:masterMosaic}) associated with rejuvenation of quiescent
stellar systems \citep{shapiro2010}, consistent with a frosting interpretation.

Both the global and core colours of the dust-ring pE galaxy (nyu674881)
reside in the gap between star-forming and quiescent galaxies. Its
relatively flat colour gradient is consistent with local quiescent
ETGs; the dust ring appears to have little effect on the its colour
gradient in contrast to the other two galaxies in this region. Yet,
previous analysis of elliptical galaxies with prominent optical dust
lanes (rings) indicates that these features represent a transition
phase between starburst and quiescence and implicate recent gas-rich
minor (i.e., ${\rm M_{1,*}}:{\rm M_{2,*}}\gtrsim4$) merging as their
likely cause \citep{shabala2012}. The capture of a gas-rich satellite
via minor merging has been shown to induce low-level star formation
consistent with a frosting interpretation \citep{kaviraj2009}. We
note that if minor merging is responsible for the enhanced star formation,
the galaxy's generally smooth morphology indicates that the star formation
lifetime appears to be greater than the time for the interaction features
to disappear. Overall, all three galaxies from region 2 of the Balmer-${\rm D_{n}(4000)}$
plane lack the clearly young spectral signatures and blue core colours
seen in the more disturbed SPM and pE/SPM galaxies and are hence not
as likely to be remnants of recent major mergers.

\subsection{Many Paths into the Green Valley?}

\label{sub:formationHistory_manyPaths}The five galaxies that principally
lie in the third region of the Balmer-${\rm D_{n}(4000)}$ plane exhibit
a tight clustering of both radial indices with a generally negative
(i.e., smaller values at larger radii) slope in their 4000\AA\  break
strengths and a positive slope in their Balmer absorption values (Fig.
\ref{fig:indicesByRegion}). The central ${\rm D_{n}(4000)}$ index
values in these galaxies are the largest seen at any radius of any
galaxy in our sample suggesting that these galaxies are the least
likely to be remnants of \textit{recent} gas-rich major mergers due
to old ($\gtrsim5\,{\rm Gyrs}$) populations in their cores. The overall
decreasing 4000\AA\ break strength and increasing (but still negative)
Balmer absorption at large radii are consistent with population gradients
seen in the general population of nearby ETGs \citep{gonzalezdelgado2014thestar,sanchez-blazquez2006stellar}.
The anomalous large, positive Balmer absorption value seen at the
sixth annulus of nyu619567 is likely a measurement error stemming
from an over-fitting of the ${\rm H\delta_{A}}$ absorption feature
(see Fig. \ref{fig:emSubSpectra}).

These galaxies have quiescent global and core colours with small,
negative gradients (i.e., slightly redder core colours, Fig. \ref{fig:colorGradients})
consistent with the locus of local quiescent ETGs. Their stellar masses%
\footnote{Using $h=0.7$.%
} are in excess of $3\times10^{10}{\rm M_{\odot}}$-- the characteristic
mass above which local galaxies tend to have little star formation
and increasingly become bulge-dominated ellipticals \citep{kauffmann20032}.
These galaxies have a median scatter in their (\textit{g-r}) colours
blue-ward from the empirical red/blue dividing line in Fig. \ref{fig:sampleSelection_colorMass}
of $\sim0.03\,{\rm mags}$. This is within the typical colour error
($0.04\,{\rm mags}$) found by M14 and may account for their apparent
green-valley colours. It is also possible that these galaxies, like
those found in region 2, may be red sequence galaxies which have migrated
into the green valley. Notably, however, they lack the Balmer absorption
seen in the region 2 galaxies and have luminosity-weighted ages at
least $1-2\;{\rm Gyrs}$ older.

The galaxies in region 3 present a rather homogeneous group reminiscent
of local relaxed ellipticals with smooth cores, quiescent colours,
steep index gradients, and no dust features. Further, all of the elliptical
(i.e., not SPM) galaxies in our sample classified as pE owing to outer
asymmetries (e.g., nyu916578) lie in this region. This is surprising
as we argued previously that the galaxies in region 2 are likely to
be remnants of minor mergers, but those galaxies lack the structural
signatures of recent interaction (i.e., arms, loops, shells, etc.)
seen in these pE galaxies. This raises the important question as to
how the galaxies in region 3 could plausibly come about.

It is well known that mergers between equal-mass disk galaxies result
in remnants with tidal tails and plumes \citep{toomre1972,barnes1988,barnes1992},
but the low surface brightness of these and other late-stage features
(e.g., small loops, arms, and shells) are also seen in minor mergers
\citep{feldmann2008} making discriminating between these two types
of merger remnants difficult \citep{lotz2008,struck2012}. Further,
not all morphologically disturbed ellipticals are star-forming \citep{michard2004},
and not all blue (presumably star-forming) ellipticals are morphologically
disturbed (see nyu22318 in our sample). Yet both plausible formation
histories predict the low-level star formation observed in nearby
ETGs \citep[e.g., ][]{yi2005}. Additionally, a scenario in which
an existing elliptical galaxy undergoes a (nearly) equal-mass gas-rich
merger would leave behind a remnant with such late-stage features.
These so-called ``mixed'' mergers have been shown to be important
in explaining the dichotomy of local elliptical galaxies \citep{rothberg20062}.
However, their general importance in galaxy evolution is not well
constrained \citep{khochfar2003}.

The modestly disturbed morphologies and quiescent colours of the galaxies
in region 3 may be described by yet another merger scenario: dissipationless
(so-called ``dry'') major mergers. In this scenario, two gas-poor
elliptical galaxies merge, producing a remnant with broad, low surface
brightness features \citep{bell2006} like those seen in many of the
pE galaxies in our sample. Assuming the progenitor galaxies posses
colours slightly redder than our empirical red/blue dividing line,
such a dry merger would conserve colour but allow the remnant to cross
our dividing line; producing a slightly bluer-than-usual elliptical
galaxy at a given stellar mass \citep{skelton2009} However, simulations
from \citet{kawata2006arered} show that the colour and structure
of such features can also be produced through spheroid-disk gas-rich
\textit{minor} mergers. The same complications listed above for mixed-mergers
then apply to this scenario, as well. \citet{sanchez-blazquez2009}
found that half of the ellipticals from \citet{vandokkum2005} having
strong tidal features exhibit a ``frosting'' of star formation like
that seen in our sample of galaxies in region 2. These frosted ellipticals
were also found to be supported by rotation, indicating the presence
of a dynamically cold stellar component. This is in contrast to simulations
of dry mergers which produce remnants with dispersion-dominated kinematics
similar to those of local cluster galaxies \citep{naab2006,naab20092}.
While our sample of galaxies in region 3 is inconsistent with the
frosting scenario, it follows that a careful analysis of the kinematic
structure of these galaxies could be used to plausibly identify if
dry mergers may play a role in their assembly.

It is clear, then, that there are many routes a galaxy may take to
arrive in region 3 of the Balmer-${\rm D_{n}\left(4000\right)}$ plane.
Extracting the precise star formation history for these galaxies requires
analysis outside the scope of this work. However, we note that in
Fig. \ref{fig:indicesByRegion}, the galaxy nyu916578 stands out among
the others in region 3 in that its three outermost radial indices
track into the fourth region of the Balmer-${\rm D_{n}(4000)}$ plane
near the burst models with luminosity-weighted ages of $\sim1\;{\rm Gyr}$
and exhibit Balmer absorption consistent with an increased fraction
of young stars. This is highly contrasted by its core population with
ages closer to $\sim10\;{\rm Gyrs}$. Such population gradients have
been seen in ``revitalized'' elliptical galaxies consistent with
ongoing star formation in their outer annuli associated with tidal
stripping during a gas-rich \textit{minor} merger \citep{fang2012theslow}.
The presence of younger stellar populations at large galactic radii
is consistent with the outside-in formation scenario of elliptical
galaxies and offers a possible explanation for observed \textquotedbl{}inverted\textquotedbl{}
age gradients and lower metallicity at large radii in some local elliptical
galaxies \citep[e.g., ][]{baes2007,labarbera2010}.

\section{Discussion}

\begin{figure*}
\begin{centering}
\includegraphics{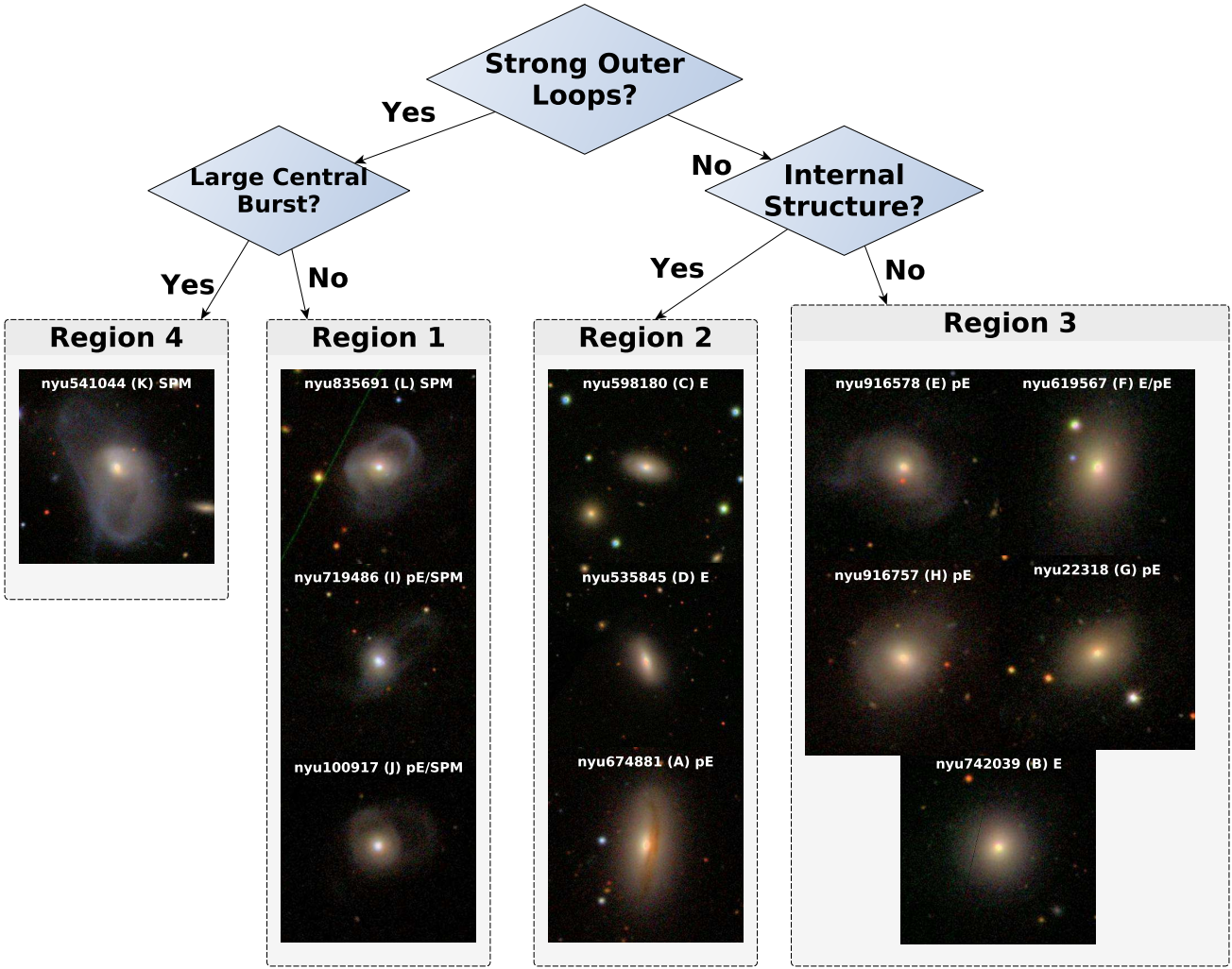}
\par\end{centering}

\protect\caption{\label{fig:decisionTree}A characteristic ``decision'' tree for
the classification of our sample of massive, blue spheroidal galaxies.
The images come from the SDSS and are scaled to $40\,{\rm kpc/h}$
on a side. At the top of each image, we list the name, single-letter
designation, and morphological classification from Table \ref{tab:sample}.
The ${\rm T_{ppm}}$ and morphological characteristics are detailed
in Section \ref{sub:sampleSelection}. Each large box represents one
of the regions of the Balmer-${\rm D_{n}\left(4000\right)}$ plane
as described in Section \ref{sub:indexPlane}.}
\end{figure*}

Using the predictions of the modern merger hypothesis as a guide,
we measure the radial Balmer and ${\rm D_{n}\left(4000\right)}$ indices
to attempt to distinguish between galaxies originating from a recent,
gas-rich major merger and those undergoing some other mass assembly
mechanism. In Section \ref{sec:dissectingFormationHistories}, we
present a detailed analysis combining the radial indices, optical
colours, and morphological features to examine the star formation
histories of our sample of galaxies. With those details in mind, we
show a characteristic ``decision'' tree (Fig. \ref{fig:decisionTree})
for our sample which allows us to \textit{a posteriori} correlate
the characteristics of each galaxy in our sample with its location
in the qualitatively defined regions of the Balmer-${\rm D_{n}\left(4000\right)}$
plane.

We begin by splitting our sample into two groups using the presence
of strong outer loops. This cleanly separates strongly disturbed spheroidal
post merger (SPM) signatures (i.e., ${\rm T_{ppm}\ge10}$) from ellipticals
with smooth or modestly disturbed morphologies. For the most disturbed
galaxies, we further examine whether they are consistent with undergoing
a strong central burst of star formation. As discussed in Section
\ref{sub:formationHistory_starburst}, the key distinction between
galaxies residing in regions 1 and 4 appears to be only the strength
of the burst and not the \textit{existence} of a burst. Further, we
find that galaxies in region 4 appear to be consistent with E+A or
post-starburst galaxies. However, we note that our small sample statistics
preclude extending this result further. Galaxies in region 1 of the
index plane are consistent with a dichotomy of star formation histories
from long-lived, continuous star formation to a recent, weak burst,
forming $\le5\%$ of their mass in a recent burst.

For the second half of the tree, we find that galaxies with smooth
elliptical morphologies but internal structures such as circumnuclear
blue rings or dust lanes preferentially reside in region 2 of the
index plane. Additionally, they possess flat gradients in their 4000\AA\ 
break strengths, but have radial Balmer absorption indices consistent
with a changing fraction of young stars atop a uniform, older population.
As outlined in Section \ref{sub:formationHistory_dusty}, previous
examination of galaxies with similar internal structures attribute
these features to revitalized stellar populations brought on by minor
merger events- in agreement with our frosting interpretation. This
is in accord with the general idea that minor mergers are thought
to play a significant role in evolving spheroidal galaxies for two
reasons: (i) the small number of major mergers in the local universe
cannot provide the abundance of green-valley ETGs \citep{darg2010,salim2010starformation,lotz2011},
and (ii) the observed size evolution of elliptical galaxies since
${\rm z\sim1}$ is best explained by growth dominated by minor mergers
\citep{bedorf2013theeffect,bezanson2009,hilz2013,lee2013,naab2009,oser2012}.

Moving to the final branch of the decision tree, we anticipate finding
smooth (i.e., small ${\rm T_{ppm}}$ values) galaxies with star formation
histories least like those predicted in the modern merger hypothesis.
Indeed, all but one of the galaxies in region 3 have star formation
histories much like local quiescent ETGs. However, many of them exhibit
external structures such as arms or loops, and all of the elliptical
(i.e., \textit{not} SPM) galaxies labelled as pE owing to these outer
asymmetries (e.g., nyu916578) lie in this region. The one galaxy which
runs counter to this general interpretation is nyu916578 which exhibits
a strong, negative age gradient with stellar populations at its largest
radii having luminosity-weighted ages $\lesssim1\,{\rm Gyr}$. Such
``inverted'' age gradients are attributable to gas-rich minor merging.
Overall, the galaxies in region 3 have a scatter in their (\textit{\small{}g-r})
colours of only $\sim0.03$ mags blue-ward of the empirical red/blue
dividing line (Fig. \ref{fig:sampleSelection_colorMass}) and lack
clear indications of ongoing star formation. It is likely that their
green valley colours will quickly fade and migrate them onto the red
sequence. Additionally, they possess stellar masses near $\sim10^{10.8}{\rm M_{\odot}h^{-2}}$
and extinction- and k-corrected SDSS absolute r-band magnitudes between
$-21\le{\rm M_{r}}\le-22$ (Table \ref{tab:sample}), placing them
close to the knee of the green valley luminosity function found by
\citet{gonccalves2012quenching}, consistent with the downsizing of
the red sequence (i.e., the low-mass end of the red sequence is populated
at late times).

Previous detailed examination of green valley galaxies revealed a
dichotomy in their morphologies \citep{schawinski2014thegreen}. The
presence of both early- and late-type galaxies in the green valley
lends support to the idea that multiple mass-assembly mechanisms are
likely responsible for galaxy growth over cosmic time. Our targeted
study of spheroidal galaxies with blue sequence colours expands on
that work, demonstrating a further branching of mass assembly histories
even within just early-type galaxies in the green valley. This indicates
that the green valley is a rich laboratory for studying the transitory
phases of galaxy formation and evolution. Our analysis could only
be accomplished via the usage of spatially-resolved spectroscopy and
motivates the study of green valley galaxies in large IFU surveys.

\section{Summary and Conclusions}

In this work, we have combined visual morphologies with spatially-resolved
spectra and optical colours to characterize the radial SFHs of a unique
sample of plausible new ellipticals. Utilizing techniques for examining
SFHs previously used to characterize the central regions of local
galaxies (e.g., K03), we compare the stellar age indices ${\rm H\,\delta}$,
${\rm H\,\gamma}$, and the 4000\AA\ break to synthetic SFH models
at radii from the core out to $\sim2-3$ half-light radii. We find
that these index values and their radial dependence correlate with
specific morphological features such that the most disturbed galaxies
have the smallest 4000\AA\  break strengths and the largest Balmer
absorption values. With this new spatial data, we are able to detect
the presence of a centrally concentrated starburst occurring within
the last $\sim1\,{\rm Gyr}$ -- allowing us to plausibly discriminate
between the construction of \textit{new} ellipticals by gas-rich major
merging following the modern merger hypothesis \citep{hopkins2008}
and ellipticals undergoing other mass assembly mechanisms.

Our main conclusions are as follows.

\renewcommand{\labelenumi}{(\roman{enumi})}
\begin{enumerate}
\item We find that one-third of our sample possess the strong morphological
disturbances and central star formation consistent with being plausible
remnants of a recent, gas-rich major merger. However, only one of
them exhibits the characteristics of having recently undergone a strong,
centrally-concentrated burst of star formation followed by quenching.
Its quiescent optical colours and strong burst signature are consistent
with its well-known status as an E+A (post-starburst) galaxy. Assuming
our small sample comports with the population of local merger remnants,
we infer that the modern merger hypothesis describes the assembly
of only $\sim25\%$ of plausible major merger remnants in the local
universe.
\item Three of the four plausible major mergers in our sample have star-forming
optical colours with much bluer centres. Their Balmer absorption and
4000\AA\  break strength index values indicative of having centrally-concentrated
young stellar populations. However, they lack the clear signatures
of a recent, \textit{strong} central burst of star formation. Overall,
their star formation histories are generally consistent with having
formed $\sim5\%$ of their stellar mass in a recent, weak burst.
\item Two-thirds of our sample of 12 massive (${\rm M_{*}}\ge10^{10}{\rm M_{\odot}}$),
nearby (${\rm z}\le0.03$), visually-selected plausible new ellipticals
with blue-sequence optical colours and varying degrees of morphological
peculiarities are inconsistent with the predictions from simulations
of gas-rich major mergers. We therefore find that their mass assembly
histories are not well-described by the modern merger hypothesis.
Half of these galaxies exhibit characteristics seen in remnants of
recent, gas-rich \textit{minor} mergers and the other half have radial
star formation histories consistent with those seen in local quiescent
ellipticals.
\end{enumerate}
We are happy to thank Hans-Walter Rix, Anna Gallazzi, and Enrique
P\'erez for useful insights and discussions. T.H. acknowledges support
from the Missouri Consortium of NASA's National Space Grant College
and Fellowship Program. D.H.M. acknowledges support from the Research
Corporation for Science Advancement under the Cottrell College Science
Award grant No. 10777. Funding for the SDSS has been provided by the
Alfred P. Sloan Foundation, the Participating Institutions, the National
Aeronautics and Space Administration, the National Science Foundation,
the U.S. Department of Energy, the Japanese Monbukagakusho, and the
Max Planck Society. The SDSS Web site is http://www.sdss.org/. The
SDSS is managed by the Astrophysical Research Consortium (ARC) for
the Participating Institutions, which are The University of Chicago,
Fermilab, the Institute for Advanced Study, the Japan Participation
Group, The Johns Hopkins University, Los Alamos National Laboratory,
the Max-Planck-Institute for Astronomy (MPIA), the Max-Planck-Institute
for Astrophysics (MPA), New Mexico State University, University of
Pittsburgh, Princeton University, the United States Naval Observatory,
and the University of Washington. This research is (partially) based
on data from the MILES project. This publication also made use of
NASA's Astrophysics Data System Bibliographic Services. Based on observations
collected at the Centro Astronómico Hispano Alemán (CAHA) at Calar
Alto, operated jointly by the Max-Planck Institut für Astronomie and
the Instituto de Astrofísica de Andalucía (CSIC). This paper is (partially)
based on data obtained by the CALIFA Survey, funded by the Spanish
Ministry of Science under grant ICTS-2009-10 {[}and add-ons{]}, and
the Centro Astronómico Hispano-Alemán.\bibliographystyle{mn2e}
\bibliography{massiveBlueEllipticals_I}

\appendix

\section{Constructing Burst Models}

\label{sec:app_constructingBurstModels}
\begin{figure}
\begin{centering}
\includegraphics{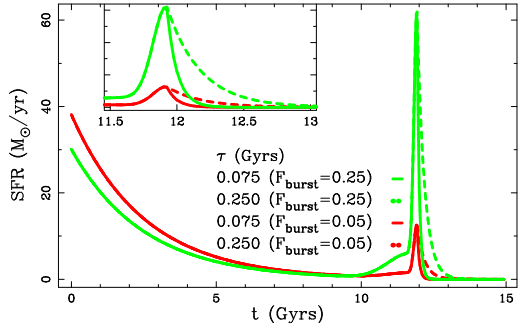}
\par\end{centering}

\protect\caption{\label{fig:burstSFHs}Star formation rates as a function of time (i.e.,
the star formation histories) for four of our constructed burst models
used throughout the rest of this work: two in which 25\% of the total
galaxy mass is formed in a single burst (green) and two burst models
with a 5\% burst fraction (red). The solid lines refer to short e-folding
times of $75\unit{{\rm \, Myrs}}$ and the dashed lines to long e-folding
times of $250\unit{{\rm \, Myrs}}$. The bursts are triggered at redshift
$0.12$ when the galaxy is $\sim12\unit{{\rm \, Gyrs}}$ old. Inset:
zoom-in of star formation histories near the burst.}
\end{figure}

A key prediction of the modern
merger hypothesis is that merging galaxies with favourable gas dynamics
undergo a centrally-concentrated burst of star formation. To determine
if the SFHs of our observed galaxies are consistent with this scenario,
we construct a suite of synthetic SEDs that model the abrupt change
in star formation during the merger process. We begin with a continuous
SFH model assuming an exponentially-decreasing star-formation rate
(SFR), ${\rm SFR_{bg}}\left(t\right)=\alpha_{0}e^{-t/\tau_{0}}$,
where $\alpha_{0}$ is the SFR at $t=0$ and $\tau_{0}$ is the characteristic
time. Atop this background of continuous star formation, we add Gaussians
of fixed widths and amplitudes to model the bursts during the merger
process. To make our models as realistic as possible, we use two bursts:
a small-amplitude, large width (i.e., long-lasting) burst representing
star formation induced by tidal interaction, and a second burst with
large-amplitude and small width representing the starburst induced
by coalescence (see Figure \ref{fig:burstSFHs}). The modern merger
hypothesis also predicts the ignition of an active galactic nucleus
(AGN) that provides a source of energetic feedback to quench star
formation in the galaxy. To model this AGN, we impose an exponentially-decreasing
SFR once the second burst has reached its peak to quench the star
formation.

Using $t_{i}$ for the time of tidal interaction and $t_{c}$ for
the time of coalescence, the SFR due to both bursts is given by
\begin{align*}
{\rm SFR_{b}}\left(t\right) & =\left[G_{i}+G_{c}\right]g\left(t\right)
\end{align*}
with
\begin{align*}
G_{x}\left(t\right) & =\alpha_{x}\exp\left\{ -\left(t-t_{x}\right)^{2}/2\sigma_{x}^{2}\right\} 
\end{align*}
and
\begin{align*}
g\left(t\right) & =\begin{cases}
1 & t<t_{c}\\
\exp\left\{ -t/\tau_{b}\right\}  & t\ge t_{c}
\end{cases}
\end{align*}
where $t_{x}$ is one of $t_{i}$ or $t_{c}$, $\alpha_{x}$ is the
burst amplitude at $t=t_{x}$, $\sigma_{x}$ is the width of the respective
burst, and $\tau_{b}$ is the characteristic time for the exponentially-declining
SFR of the burst at coalescence. The fraction of stars formed (by
mass) during the burst, ${\rm F_{burst}}$, is defined as
\begin{align*}
{\rm F_{burst}} & =\frac{{\rm stellar\, mass\, formed\, during\, burst}}{{\rm total\, stellar\, mass}}.
\end{align*}
Incorporating our definitions from above, this becomes
\begin{align}
{\rm F_{burst}} & =\frac{\int_{t_{0}}^{t_{f}}{\rm SFR_{b}}\left(t\right)dt}{\int_{t_{0}}^{t_{f}}{\rm SFR_{bg}\left(t\right)+SFR}_{b}\left(t\right)dt}\nonumber \\
 & =\frac{\int_{t_{0}}^{t_{f}}\left[G_{i}+G_{c}\right]g\left(t\right)dt}{\int_{t_{0}}^{t_{f}}\left[\alpha_{0}\exp\left\{ -t/\tau_{0}\right\} +G_{i}+G_{c}\right]g\left(t\right)dt}\label{eq:fburst}
\end{align}
where ${\rm SFR_{bg}}$ is the star formation rate of the background
model. We assume that this background star formation continues during
the burst such that the total number of stars formed is the background
star formation plus the two bursts.

Clearly, there are many parameters to be tuned in this formula. We
choose to keep constant the widths of the bursts and the characteristic
time of the background star formation. Specifically, we assume a total
galaxy lifetime $t_{f}=t_{c}+3\unit{Gyrs}$, a characteristic time
of $\tau_{0}=2.5\unit{Gyrs}$ for the background star formation, an
interaction time of $t_{i}=t_{c}-250\unit{Myrs}$, an interaction
burst width of $\sigma_{i}=750\unit{Myrs}$, and a coalescence burst
width of $\sigma_{c}=100\unit{Myrs}$. These parameters are consistent
with the results of gas-rich major merger simulations from \citet{cox2008}.
Finally, we normalize the total (background plus bursts) star formation
such that a mass of ${\rm M_{tot}}$ is formed.

The amplitudes of the Gaussian bursts are independent of time, so
we can rearrange the integrals in Equation \ref{eq:fburst} using
the substitutions
\begin{align*}
\gamma_{1} & =\int_{t_{0}}^{t_{f}}\exp\left\{ -\left(t-t_{i}\right)^{2}/2\sigma_{i}^{2}\right\} g\left(t\right)dt\\
\gamma_{2} & =\int_{t_{0}}^{t_{f}}\exp\left\{ -\left(t-t_{c}\right)^{2}/2\sigma_{c}^{2}\right\} g\left(t\right)dt\\
\gamma_{3} & =\int_{t_{0}}^{t_{f}}\exp\left\{ -t/\tau_{0}\right\} g\left(t\right)dt
\end{align*}
to generate a set of simple algebraic expressions to constrain the
peak SFR parameters $\alpha_{0}$, $\alpha_{c}$, and $\alpha_{i}$
such that
\begin{align*}
\alpha_{i}\gamma_{1}+\alpha_{c}\gamma_{2} & ={\rm M_{tot}}-\alpha_{0}\gamma_{3}\\
\alpha_{i}\gamma_{1}+\alpha_{c}\gamma_{2} & =\frac{{\rm F_{burst}}}{1-{\rm F_{burst}}}\alpha_{0}\gamma_{3}
\end{align*}
If we assume the peak SFR of the interaction and coalescence bursts
are tied such that $\alpha_{i}=k\alpha_{c}$, then we arrive at our
final expressions for the peak SFR parameters
\begin{align*}
\alpha_{0} & =\frac{{\rm M_{tot}}\left(1-{\rm F_{burst}}\right)}{\gamma_{3}}\\
\alpha_{c} & =\frac{{\rm M_{tot}}{\rm F_{burst}}}{\left(k\gamma_{1}+\gamma_{2}\right)}.
\end{align*}
Using this model, we parametrize the star formation history of our
burst models using four parameters: the burst fraction ${\rm F_{burst}}$,
the relative strength of the peak SFR of the two bursts $k$, the
characteristic time of the post-coalescence exponentially-decaying
SFR $\tau_{b}$, and the total mass formed. We assume $k=0.1$ and
${\rm M_{tot}=10^{11}M_{\odot}}$ throughout, but generate our suite
of burst models based on a variety of burst fractions and characteristic
times. See Section \ref{sec:modellingSFHs} for these values.

\section{Index Model Plane}

\label{sec:app_indexModelPlane}In Fig. \ref{fig:modelComparison},
we present the detailed Balmer-${\rm D_{n}\left(4000\right)}$ index
plane for each galaxy in our sample. For each galaxy, we show the
suite of models outlined in Section \ref{sec:modellingSFHs}. The
inset in each figure shows the radial order of the index measurements
with 1 indicating the central annulus.
\begin{figure*}
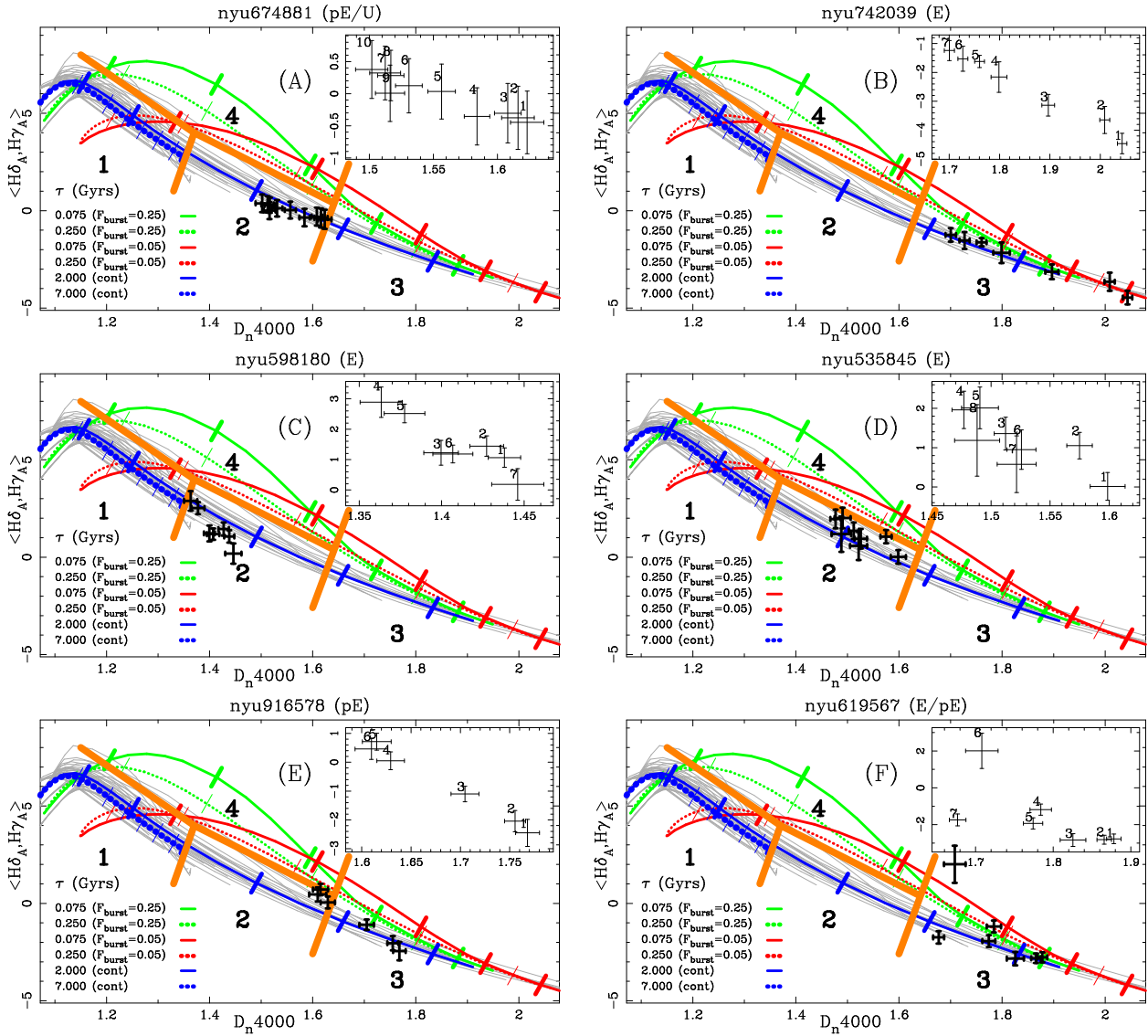

\begin{tabular}{cc}
\includegraphics{modelAging/nyu674881} & \includegraphics{modelAging/nyu742039}\tabularnewline
\includegraphics{modelAging/nyu598180} & \includegraphics{modelAging/nyu535845}\tabularnewline
\includegraphics{modelAging/nyu916578} & \includegraphics{modelAging/nyu619567}\tabularnewline
\end{tabular}

\protect\caption{\label{fig:modelComparison}The stellar Balmer absorption versus 4000\AA\
break strength model plane showing two continuous SFH models (blue)
and two burst models (green) covering a range of e-folding times permitted
for each model type. The full suite of continuous SFH models are shown
in light grey. Tick marks denote points along the model lines with
known ages. For the continuous models, the ages are 1, 3, 5, 7, 9,
and 11 Gyrs. For the burst models, the ages are 0.2, 0.5, 0.9, 1.5,
and 2.5 Gyrs. The four qualitative regions from the text are labelled
and the large orange lines mark their approximate boundaries. At the
top of each panel, we give the galaxy's morphological classification
next to its name. Inset: Zoom-in to the plane showing the indices
marked in annulus order.}
\end{figure*}
\begin{figure*}
\ContinuedFloat%
\begin{tabular}{cc}
\includegraphics{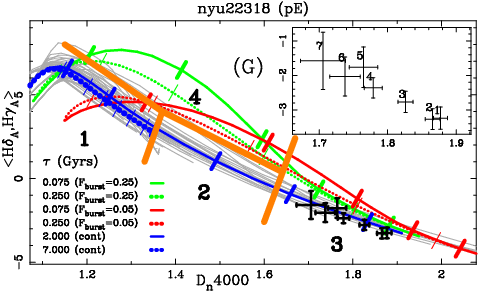} & \includegraphics{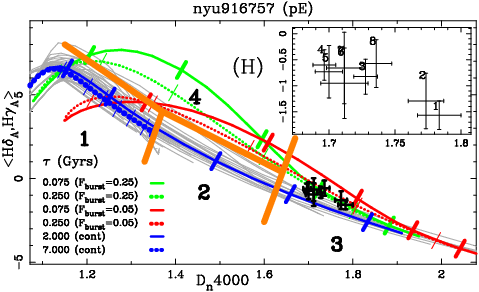}\tabularnewline
\includegraphics{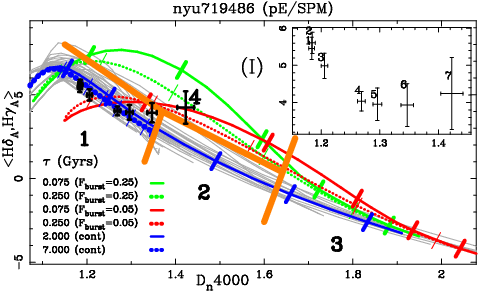} & \includegraphics{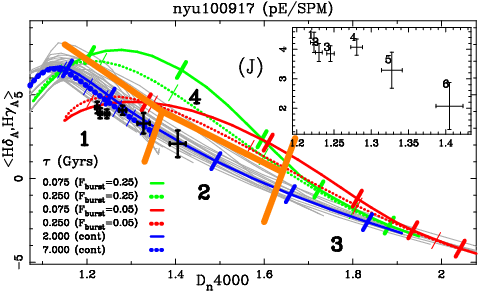}\tabularnewline
\includegraphics{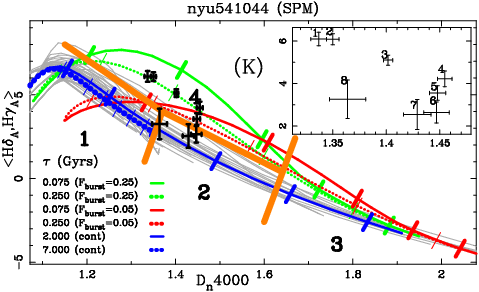} & \includegraphics{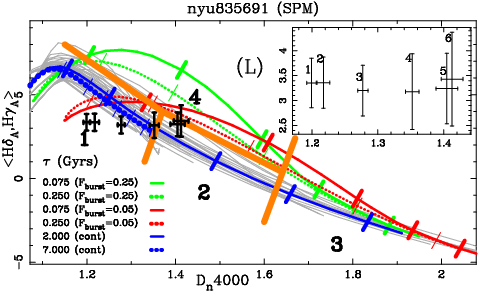}\tabularnewline
\end{tabular}

\protect\caption{(cont.)}
\end{figure*}

\section{Radial Spectra}

\label{sec:app_radialSpectra}In Fig. \ref{fig:emSubSpectra}, we
present the radially-binned, emission-subtracted spectra for each
galaxy in our sample. The number of spectra is determined by a signal-to-noise
cut-off outlined in Section \ref{sub:constructingBinnedSpectra}.
The spectra are listed top-to-bottom from the centre to the outer
most annulus. For the central spectrum, we label several well-known
spectral lines. Note, that due to the spectra being produced on different
spectrographs with varying setups (see Section \ref{sub:observations}
for details), the wavelength coverage is not the same for each galaxy.
\begin{figure*}
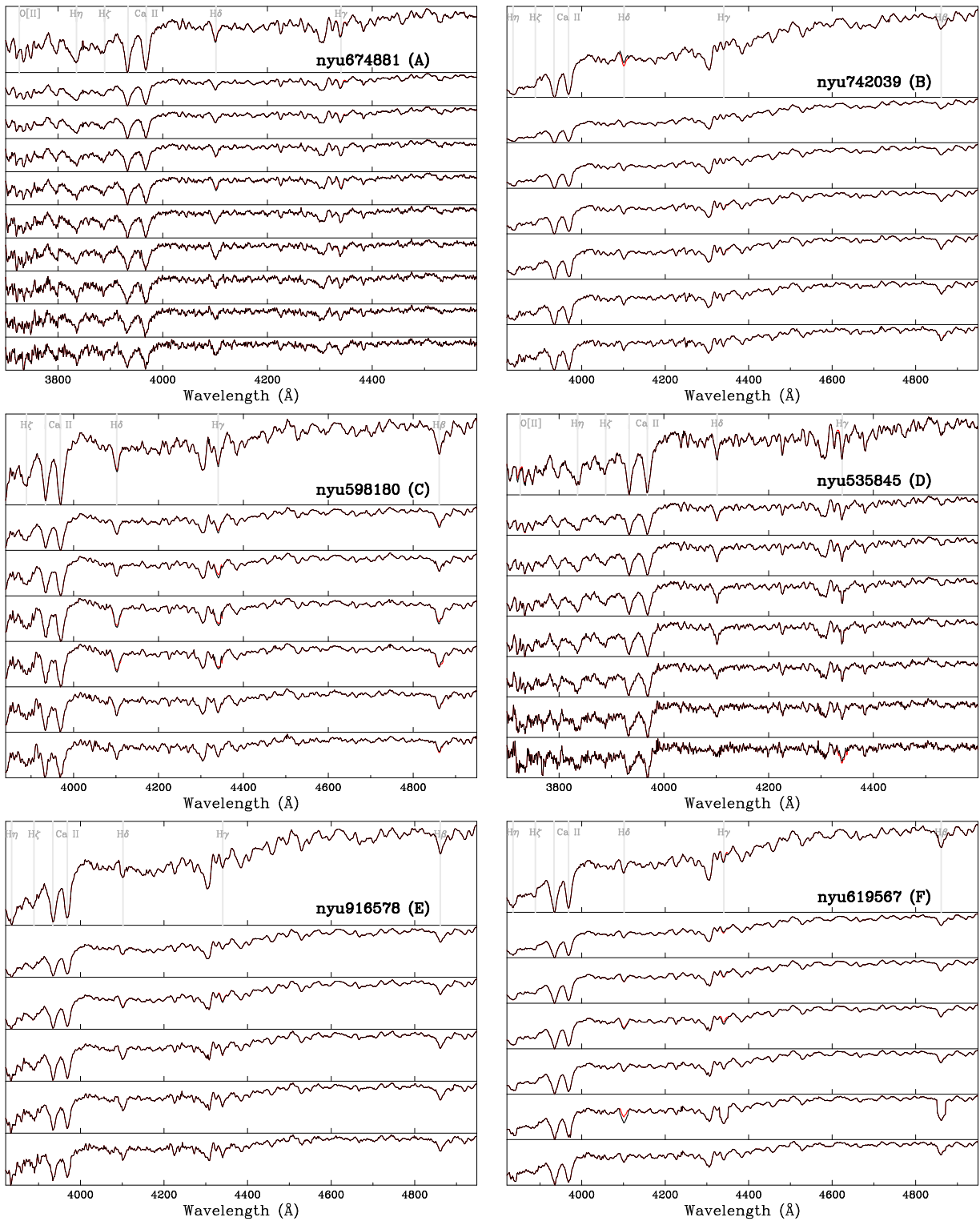

\begin{tabular}{cc}
\includegraphics{indices/nyu674881\lyxdot modelfits} & \includegraphics{indices/nyu742039\lyxdot modelfits}\tabularnewline
\includegraphics{indices/nyu598180\lyxdot modelfits} & \includegraphics{indices/nyu535845\lyxdot modelfits}\tabularnewline
\includegraphics{indices/nyu916578\lyxdot modelfits} & \includegraphics{indices/nyu619567\lyxdot modelfits}\tabularnewline
\end{tabular}

\protect\caption{\label{fig:emSubSpectra}Observed spectrum (red) and emission-subtracted
spectrum (grey) for each annulus with S/N/pixel $>10$ from each galaxy
in our sample. Each spectrum is scaled to arbitrary flux units to
show detail and given in annulus order from top to bottom with the
central annulus at the top. For the central annulus, several well-known
spectral features such as the hydrogen Balmer series and the Ca II
lines are highlighted with grey vertical lines and labelled accordingly.}
\end{figure*}
\begin{figure*}
\ContinuedFloat%
\begin{tabular}[t]{cc}
\includegraphics{indices/nyu22318\lyxdot modelfits}  & \includegraphics{indices/nyu916757\lyxdot modelfits}\tabularnewline[0.25cm]
\includegraphics{indices/nyu719486\lyxdot modelfits}  & \includegraphics{indices/nyu100917\lyxdot modelfits}\tabularnewline[0.25cm]
\includegraphics{indices/nyu541044\lyxdot modelfits}  & \includegraphics{indices/nyu835691\lyxdot modelfits}\tabularnewline[0.25cm]
\end{tabular}

\protect\caption{(cont.)}
\end{figure*}

\end{document}